\documentclass[aps, pra, a4paper, showpacs, twocolumn, english, 10pt, floatfix]{revtex4-1}
\usepackage[T1]{fontenc}
\usepackage{babel}
\usepackage{bbm, amsthm, bm, textcomp, nicefrac, geometry, ragged2e}
\usepackage[dvipsnames]{xcolor}
\usepackage{float}
\usepackage[bbgreekl]{mathbbol}
\usepackage{graphicx, epstopdf, color, verbatim, enumitem, ulem}
\usepackage{pbox, array}
\usepackage{mathrsfs}
\usepackage{amssymb}
\usepackage{mathtools}
\usepackage{stackrel}
\usepackage[thinlines]{easytable}
\usepackage{amsmath}
\makeatletter

\usepackage{array}
\usepackage{multirow}

\usepackage[caption=false]{subfig}
\addto\captionsenglish{}

\usepackage{hyperref}
\hypersetup{
    colorlinks = true,
    linkcolor = BrickRed,
    citecolor = BrickRed,
    filecolor = Black,      
    urlcolor = Black,
}

\begin{document}
\title{Noisy Stabilizer Formalism}

\author{Maria Flors Mor-Ruiz$^1$ and Wolfgang D\"ur$^1$}

\affiliation{$^1$Universit\"at Innsbruck, Institut f\"ur Theoretische Physik, Technikerstra{\ss}e 21a, 6020 Innsbruck, Austria}

\date{\today}
\begin{abstract}
Despite the exponential overhead to describe general multi-qubit quantum states and processes, efficient methods for certain state families and operations have been developed and utilized. The stabilizer formalism and the Gottesman-Knill theorem, where pure stabilizer or graph states are manipulated by Clifford operations and Pauli measurements, are prominent examples, and these states play a major role in many applications in quantum technologies. Here we develop a noisy stabilizer formalism, i.e., a method that allows one to efficiently describe and follow not only pure states under Clifford operations and Pauli measurements but also Pauli noise processes acting on such stabilizer states, including uncorrelated and correlated dephasing and single- or multiple-qubit depolarizing noise. The method scales linearly in the number of qubits of the initial state, but exponentially in the size of the target state. Thus, whenever a noisy stabilizer state is manipulated by means of local Pauli measurements such that a multipartite entangled state of a few qubits is generated, one can efficiently describe the resulting state. 
\end{abstract}

\maketitle

\section{Introduction}\label{sec:introduction}
The classical description of compound quantum systems consisting of multiple components is in general difficult due to the exponential scaling of the Hilbert space dimension with the number of systems. This provides a severe hindrance to our understanding of large-scale quantum systems, and the classical simulation of multi-system states and processes. Nevertheless, several methods to efficiently describe and simulate certain classes of quantum states, operations and measurements have been developed and utilized. Prominent examples that have found multiple applications in quantum information theory and quantum technologies are stabilizer or graph states \cite{nielsen_chuang_2010, stabilizer_gottesman, hein_multiparty_2004, hein_entanglement_2006} and their manipulation by Clifford operations and Pauli measurements. Rather than describing the state vector with its exponentially many coefficients directly, one specifies the state using its Pauli stabilizers, where only linearly many stabilizers are required. One can also update these stabilizers when performing Clifford operations and Pauli measurements, leading to the Gottesman-Knill theorem \cite{gottesman_1998, aaronson_2004, anders_2006, nest_gottesman} stating that all quantum computations consisting of such operations only can be efficiently simulated classically. Stabilizer states described by Pauli stabilizers are (up to local unitary operations) equivalent to so-called graph states \cite{hein_multiparty_2004, hein_entanglement_2006}, which have found many applications in measurement-based quantum computation \cite{briegel_measurement_based_2009, raussendorf_measurement}, quantum error correction \cite{pirker_construction_2017, schlingemann_error}, secret sharing \cite{markham_graph, bell_experimental_2014} and quantum metrology \cite{shettell_graph}.

However, in realistic situations, one also must deal with noise, imperfections and decoherence and wants to analyze the performance of protocols under such circumstances. In these situations, a description of the system in terms of pure states is not sufficient; one needs to use density matrices to take errors and noise processes into account. This not only squares the required number of parameters but also makes the description of operations, as well as their action on density matrices, more complicated and costly. Furthermore, it is often not clear how to extend a formalism that is valid for pure states to mixed states. 

Here we extend the stabilizer formalism to such realistic situations and show how to deal with specific noise processes, namely (quasi) local noise processes and imperfect measurements that are diagonal in the Pauli basis. The rules used to update the resulting state after Pauli measurements and Clifford operations can also be utilized to update Pauli noise operators that act on such stabilizer states. That is, we follow each of the noise operators after Pauli measurements and Clifford operations and obtain their effective description on the final state. This is done in addition to the update of the initial pure state. To this aim we make use of the following facts: (i) The action of any Pauli operator on a graph state can be expressed, up to a phase, in terms of commuting $\sigma_z$ operators only; (ii) any local noise process can be made diagonal in the Pauli basis and is then described by a constant number of Pauli noise operators; (iii) updates of noise operators under Clifford operations and Pauli measurements are efficient and the final noise operators only act on a system whose size is given by the number of qubits of the final state. For Pauli diagonal maps, it suffices to update the maps independently. That is, for a graph state of size $n$ that is affected by local noise, noisy Clifford operations and $n-m$ noisy Pauli measurements, one can describe the final resulting state of $m$ qubits with an overhead that is linear in the size of the initial state $n$ but exponential in $m$, the size of the target state.    

This provides us with a powerful tool to study the influence of noise and imperfections in the manipulation of graph states. To illustrate our method, we discuss in detail the example of generating Bell states from a noisy one-dimensional (1D) cluster state of arbitrary size, where each qubit is subjected to local Pauli noise. This is, for instance, a relevant noise model when considering imperfect quantum memories in systems where qubits are distributed among multiple sites, as in, e.g., a quantum network scenario. We show, perhaps surprisingly, that the fidelity of the resulting Bell state depends on the order in which the intermediate systems are measured. We also discuss other potential applications of our noisy stabilizer formalism in the context of quantum communication and computing. 

The paper is organized as follows. In Sec.~\ref{sec:background} we provide some background information on stabilizer formalism and graph states and give a brief introduction to single-qubit noise models applied to graph states. We also settle the notation we use throughout the article in this section. In Sec.~\ref{sec:noisy:stabilizer:formalism} we describe a method to efficiently describe the manipulation of noisy graph states, which we call noisy stabilizer formalism. In Sec.~\ref{sec:limits} we study the limitations of the use of this formalism for general noise models and quantum states more general than graph states. In Sec.~\ref{sec:efficiency} a discussion of the efficiency and computational application of the noisy stabilizer formalism is presented. In Sec.~\ref{sec:application} we make use of the presented method for a nontrivial example that shows explicitly the advantages of the formalism. We summarize and discuss our results in Sec.~\ref{sec:conclusion}, where we also provide an outlook on additional possible applications.

\section{Background}\label{sec:background}
In the following, we recall some basic notations and results concerning stabilizer states, graph states, and single-qubit noise models which will be used throughout the paper.
\subsection{Stabilizer Formalism}\label{ssec:stabilizer:formalism}
The $n$-qubit Pauli group $\mathcal{P}_n$ \cite{fujii_stabilizer_2015} is defined as 
\begin{equation}
    \mathcal{P}_n=\{\pm1,\pm i\}\times\{\mathbbm{1}, X, Y, Z\}^{\otimes n}.
\end{equation}
An element of the Pauli group is called a Pauli product. An $n$-qubit stabilizer group $\mathcal{S}$ \cite{fujii_stabilizer_2015} can be defined as an Abelian (commutative) subgroup of the $n$-qubit Pauli group,
\begin{equation}
    \mathcal{S} = \{S_i\} \text{ s.t.} -\mathbbm{1} \notin \mathcal{S} \text{ and } \forall S_i, S_j \in S, [S_i, S_j]=0.
\end{equation}

An element from $\mathcal{S}$ is called a stabilizer operator, and the elements in the maximally independent subset $\mathcal{S}_g$ of the stabilizer group are called the stabilizer generators. Independence in this framework means that any stabilizer generator cannot be expressed as a product of other generators. Then, any of the elements in $\mathcal{S}$ can be generated by the product of the stabilizer generators. Thus, a stabilizer group $\mathcal{S}$ can be expressed in terms of its stabilizer generators $\mathcal{S}_g$ and it is defined by $\mathcal{S}=\langle\mathcal{S}_g\rangle$. 

The stabilizer state $|{\psi}\rangle$, for a given stabilizer group $\mathcal{S}$, can be defined as a simultaneous eigenstate with eigenvalue +1 of all the stabilizer operators in $\mathcal{S}$, 
\begin{equation}
    S_i |{\psi}\rangle = |{\psi}\rangle \quad \forall S_i\in\mathcal{S}.
\end{equation}
It is sufficient if the state is an eigenstate with eigenvalue +1 of the stabilizer generators,
\begin{equation}
    g_i|{\psi}\rangle = |{\psi}\rangle \quad \forall g_i\in\mathcal{S}_g.
\end{equation}

In general, when talking about a certain stabilizer state $|{\psi}\rangle$, one can say that it is stabilized or invariant under the action of the operators in $\mathcal{S}$. Then, all the possible states that are stabilized by the subgroup $\mathcal{S}$ form $V_S$, the vector space stabilized by $\mathcal{S}$, and $\mathcal{S}$ is said to be the stabilizer of the space $V_S$.

\subsection{Clifford Operations} \label{ssec:clifford:operations}
The subset of all unitary quantum operations that map stabilizer states to stabilizer states are the so-called (local) Clifford operations \cite{nest_invariants_2005}. A Clifford operation can be defined as an operation that transforms a Pauli product into another Pauli product under its conjugation \cite{fujii_stabilizer_2015}. Consider a Clifford operation $U$ acting on a stabilizer state $|{\psi}\rangle$, defined by the stabilizer group $\mathcal{S}=\langle\{g_i\}\rangle$,
\begin{equation}
    U|{\psi}\rangle=U g_i|{\psi}\rangle=U g_iU^{\dagger}U|{\psi}\rangle=g_i' U|{\psi}\rangle,
\end{equation}
where $g_i'=U g_i U^{\dagger}$. This indicates that the state $U|{\psi}\rangle$ is stabilized by all $g_i'$. Since $U$ is a unitary Clifford operation, the group $\{g_i'\}$ is also an Abelian subgroup of the Pauli group. Therefore, the state $U|{\psi}\rangle$ is stabilized by the stabilizer group $\{g_i'\}$. 

\subsection{Graph states}\label{ssec:graph:states}
Graph states \cite{hein_multiparty_2004, hein_entanglement_2006} are a subclass of multiqubit states. These can be represented as graphs $G = (V, E)$, where $V$ denotes a finite set containing the vertices and $E$ is a set whose elements are the edges between two vertices. The state associated with this graph $G$ corresponds to the unique $+1$ eigenstate of the stabilizers
\begin{equation}\label{eq:graph:state}
    K_{a}=X_a Z_{N_a}:=X_a\prod_{b\in N_a} Z_b
\end{equation}
for all $a\in V$, where $N_a$ denotes the neighborhood, which is the set of vertices adjacent to a given vertex, of $a$. In this work, we will use the notation where $Z_{N_a}$ corresponds to the tensor product of Pauli $Z$ operators on all the qubits in $N_a$, $Z_{N_a}:=\prod_{b\in N_a} Z_b$.

Graph states can be manipulated and transformed by certain quantum operations \cite{hein_multiparty_2004, hein_entanglement_2006}. The most relevant and widely used properties are presented next.

\subsubsection{Local complementation} \label{sssec:local:complementation}
Given a vertex $a$ in the graph $G$, a local complementation $\tau_a$ acts by inverting the edges connecting the neighbors of $a$, $G'=\tau_a(G)$ such that the new neighborhood of any qubit $b\in N_a$ is $N'_b=\left(N_b\cup N_a\right) \setminus \left(N_b\cap N_a\right)\setminus\{b\}$. The graph state after this operation is 
\begin{equation} \label{eq:local:complementation}
    |{G'}\rangle = |{\tau_a (G)}\rangle = U_a^{\tau} |{G}\rangle = \sqrt{-iX_a}\prod_{b\in N_a}\sqrt{i Z_b}|{G}\rangle, 
\end{equation}
where $U_a^{\tau}$ is a local Clifford (LC) unitary. Two graph states are said to be LC equivalent if the corresponding graphs are related by a sequence of local complementations.

\subsubsection{Local Pauli measurements on graph states} 
Let $a$ be the vertex corresponding to the qubit to be measured in a Pauli basis. Corresponding to this measurement, the following unitaries are defined
\begin{align}
  U_{z,+}^{(a)} & = 1, & U_{z,-}^{(a)}&=\prod_{b\in N_a} Z_b, \label{eq:correction:z}\\
  U_{y,+}^{(a)} & = \prod_{b\in N_{a}}(-i Z_{b})^{1/2}, & U_{y,-}^{(a)} & = \prod_{b\in N_a}(i Z_b)^{1/2}, \label{eq:correction:y}
\end{align}
and, depending furthermore on a vertex $b_0\in N_a$,
\begin{equation} \label{eq:correction:x}
    \begin{aligned}
     U_{x,+}^{(a)}&= (i Y_{b_0})^{1/2}\prod_{b\in N_a-N_{b_0}-\{b_0\}}Z_b,\\ U_{x,-}^{(a)} &=(-i Y_{b_0})^{1/2}\prod_{b\in N_{b_0}-N_a-\{a\}}Z_b.
    \end{aligned}
\end{equation}

Now, the resulting state vector after the measurement of qubit $a$ in a Pauli basis, depending on the outcome $\pm 1$, is given by
\begin{equation}\label{eq:pauli:measurement}
    P_{i,\pm}^{(a)}|{G}\rangle = |{i,\pm}\rangle^{(a)}\otimes U_{i,\pm}^{(a)}|{G'}\rangle, \quad i=x, y, z,
\end{equation}
where $|{i,\pm}\rangle^{(a)}$ is the state of qubit $a$ after the measurement. This state corresponds to the eigenvector with eigenvalue $\pm 1$, depending on the measurement outcome, for the corresponding Pauli basis $i=x,y,z$. The resulting graph state $G'$ is given by
\begin{align} \label{eq:graph:z:y}
    G' = 
    \begin{cases}
    G-\{a\} \text{ for } Z, \\
    G-E(N_a,N_a) \text{ for } Y,
    \end{cases}
\end{align}
and for $X$,
\begin{equation}\label{eq:graph:x}
\begin{aligned}
    G' = & G\Delta E(N_{b_0},N-a)\Delta E(N_{b_0} \cap N_a, N_{b_0} \cap N_a) \\
    & \times \Delta E(\{b_0\}, N_a - \{b_0\}), 
    \end{aligned}
\end{equation}
where, $E(A, B)=\{\{a,b\}\in E: a\in A, b\in B, a\neq b\}$, and $E\Delta F=(E\cup F)-(E\cap F)$. Essentially the resulting graph after a $Z$ measurement is the deletion of all edges incident to the vertex $a$. For the $Y$ measurement, the neighborhood of $a$ is inverted and then all edges incident to $a$ are deleted. Finally, for the $X$ measurement, the local unitary depends on the choice of $b_0$. However, the resulting graph states arising from different choices of $b_0$ and $b'_0$ will be equivalent up to the local unitary $U_{b'_0}^{\tau}U_{b_0}^{\tau}$. Moreover, the resulting graph is the result of first inverting the neighborhood of $b_0$ and then the same for $a$, followed by deleting the edges incident on $a$, and finally inverting the neighborhood of $b_0$. The rules for the $X$ and $Y$ measurement rules can also be expressed as
\begin{align}
P_{y, \pm}^{(a)}&=(U_{a}^{\tau})^{\dagger}P_{z, \pm}^{(a)}U_{a}^{\tau}, \label{eq:trick:y} \\
P_{x, \pm}^{(a)}&=U_{b_0}^{\tau}P_{y, \pm}^{(a)}(U_{b_0}^{\tau})^{\dagger}. \label{eq:trick:x}
\end{align}

The resulting graph state after a Pauli measurement is up to a certain unitary that depends on the outcome of the measurement, defined in Eqs.~\eqref{eq:correction:z}-\eqref{eq:correction:x}. For a sequence of local Pauli measurements, the local unitaries have to be taken into account if one of the qubits to be measured is affected by the unitary. For convenience, let us define the manipulation operator for a local Pauli measurement on $a$,
\begin{equation} \label{eq:measured:graph}
    L_{i,\pm}^{(a)}=\langle{i,\pm}|^{(a)}\otimes (U_{i,\pm}^{(a)})^{\dagger}P_{i,\pm}^{(a)},
\end{equation}
where $i=x,y,z$ denotes the basis of the measurement. The action of such an operator is $|{G'}\rangle = L_{i,\pm}^{(a)}|{G}\rangle$.

\subsubsection{Merging graph states} \label{sssec:merging}
\begin{figure}[ht]
  \includegraphics[width=\columnwidth]{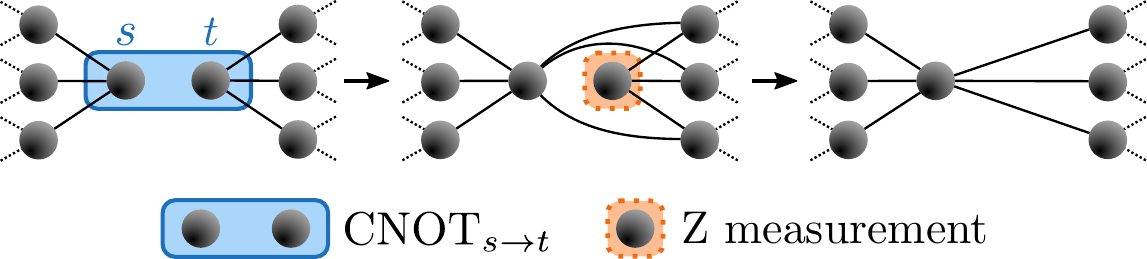}
  \caption{Graphical illustration of the merging procedure.}
  \label{fig:merge}
\end{figure}

To merge two graph states \cite{pirker_modular_2018, kruszynska_2006}, a controlled-NOT (CNOT) gate is applied between two vertices, each from a different graph. One of the qubits is denoted by $s$, meaning it is the source, and the other one by $t$, meaning it is the target. This operation introduces new edges in the resulting graph state between the source qubit and the neighborhood of the target qubit. Moreover, if the target qubit of the CNOT gate is measured in the $Z$ basis, the neighborhood of the target qubit is moved to the neighborhood source qubit. Therefore, the merging operation is defined by the operator $L_{z,\pm}^{(t)}\text{CNOT}_{s\rightarrow t}$. Figure~\ref{fig:merge} depicts the action of the merging procedure.

\begin{figure}[ht]
  \includegraphics[width=\columnwidth]{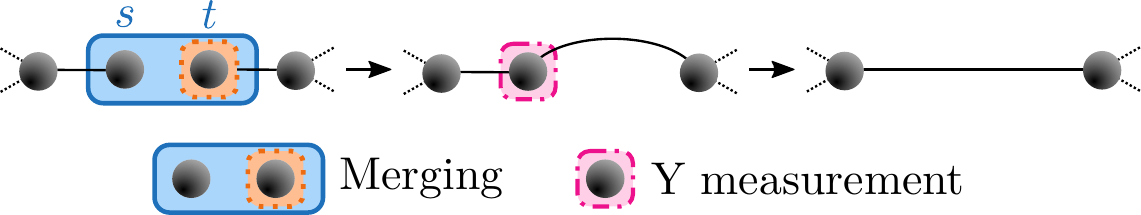}
  \caption{Graphical illustration of the full merging procedure.}
  \label{fig:full:merge}
\end{figure}

Furthermore, in order to not leave any of the merging qubits in the final graph state, one can measure $s$ in the $Y$ basis. This manipulation, defined by $L_{y,\pm}^{(s)} L_{z,\pm}^{(t)} \text{CNOT}_{s\rightarrow t}$, is referred to as the full merging procedure, whose action is graphically represented in Fig.~\ref{fig:full:merge}. Notice that this procedure is very similar to a Bell measurement \cite{nielsen_chuang_2010}. In the latter, the source qubit is measured in the $X$ basis. However, from Eq.~\eqref{eq:trick:x} one can see that the resulting graph state from the full merging is LC-equivalent to the one resulting from the Bell measurement.

\subsection{Action of noise maps on graph states}

\subsubsection{Single-qubit Pauli noise channels}
A single-qubit Pauli noise channel is described as 
\begin{equation}\label{eq:pauli:channel}
    \mathcal{E}_a\rho=\lambda_0 \mathbbm{1}_a\rho \mathbbm{1}_a + \lambda_1 X_a\rho X_a+ \lambda_2 Y_a\rho Y_a + \lambda_3 Z_a\rho Z_a,
\end{equation}
where $\sum_i\lambda_i=1$. Consider that each qubit in an $n$-qubit graph state is subject to a Pauli noise channel. Then, Eq.~\eqref{eq:pauli:channel} is applied for all $a\in V$ and $\rho=|G\rangle\langle G|$ defines the density matrix of the $n$-qubit graph state. The total noisy state is $\mathcal{E}_1\mathcal{E}_2\cdots \mathcal{E}_n \rho$.

A depolarizing channel is a particular instance of the single-qubit Pauli channel, where with probability $p$ the state remains unchanged, while with $1-p$ the qubit is depolarized, meaning that the qubit has the completely mixed state $\mathbbm{1}/2$. Thus, $\lambda_0 = (1+3p)/4$ and $\lambda_1=\lambda_2=\lambda_3=(1-p)/4$. Importantly, in \cite{dur_stability_2004} it was shown that a depolarizing channel acting on a graph state can be equivalently described by a map $\mathcal{M}_a$ whose Kraus operators only contain products of $Z$ and $\mathbbm{1}$, where $Z$ may act on qubit $a$ and its neighborhood. This follows from the fact that if one considers a graph state $|G\rangle$ and an $X$ error on qubit $a$, then
\begin{equation}
    X_a|G\rangle = X_a K_a|G\rangle = Z_{N_a}|G\rangle.
\end{equation}
Thus, an $X$ error on a certain qubit in a graph state can be translated to $Z$ errors on its neighbors. Analogously, a $Y$ error on $a$ is such that
\begin{equation}
    Y_a|G\rangle = -i Z_a Z_{N_a}|G\rangle,
\end{equation}
so that it translates into a $Z$ error on $a$ and its neighbors with a $-i$ phase.

This can be further generalized for a Pauli noise channel such that it can be rewritten as a map with Pauli noise operators of the form $Z_a^{\alpha}Z_{N_a}^{\beta}$, where $\alpha$ and $\beta$ can take values 0 and 1. Therefore, the noise channel described in Eq.~\eqref{eq:pauli:channel} is equivalent to the noise map
\begin{equation}\label{eq:pauli:map}
\mathcal{M}_a\rho=\sum_{i=0}^{3}\lambda_i\left(Z_a^{\alpha(i)}Z_{N_a}^{\beta(i)}\right)\rho\left(Z_a^{\alpha(i)}Z_{N_a}^{\beta(i)}\right),
\end{equation}
where $\alpha(0)=\alpha(1)=\beta(0)=\beta(3)=0$ and $\alpha(2)=\alpha(3)=\beta(1)=\beta(2)=1$. For such Pauli channels, the order in which the maps are applied to the graph state is irrelevant and thus they can be studied separately.

\subsubsection{Multi-qubit Pauli channels}
Similarly, one can consider multiqubit Pauli channels acting on a graph state $|G\rangle$,
\begin{equation}
    \mathcal{E}_{a_1a_2 \ldots a_k} \rho = \sum_{j=1}^{4^k} \lambda_{j} N_j \rho N_j^\dagger.
\end{equation}
Noise operators $N_j$ are all combinations of tensor products of Pauli operators, $N_j= \sigma_{j_1}^{(a_1)}\otimes\sigma_{j_2}^{(a_2)} \cdots \sigma_{j_k}^{(a_k)}$, where only diagonal terms appear. We use the shorthand notation $\sigma_0=\mathbbm{1}, \sigma_1=X, \sigma_2=Y, \sigma_3=Z$. Note that any channel can be brought to a Pauli diagonal form without changing the diagonal elements by a depolarization procedure, i.e., by applying random unitary operations before and after the application of the channel \cite{dur_standard}. 

For such Pauli diagonal channels, one can replace for each qubit the operators $X_a$ by $Z_{N_a}$ and $Y_a$ by $Z_a Z_{N_a}$. Note that phases $\pm i$ do not matter for diagonal terms $N_i \rho N_i^\dagger$. 

Note that some care is required when considering general noise channels that also include off-diagonal terms when written in the Pauli basis, as phases that appear from the $Y$ noise operators matter and additional phases due to Pauli commutation relations occur. Then a simple replacement is no longer sufficient, as we discuss in more detail in Appendix~\ref{a:remarks:general:noise}. 

\section{Noisy Stabilizer Formalism} \label{sec:noisy:stabilizer:formalism}
Our goal is to find an efficient method to describe the manipulation of noisy graph states. The straightforward approach would be to take the noisy $n$-qubit graph state $\mathcal{E}_1\mathcal{E}_2\cdots \mathcal{E}_n \rho$ and apply the operators describing manipulations of graph states, such as the ones presented in Sec.~\ref{ssec:graph:states}. Nevertheless, this is rather hard to compute for relatively large systems as we need to keep track of and operate on a $2^n\times 2^n$ density matrix. For Pauli channels, the density operators are reduced to matrices that are diagonal in the graph state basis, however, one still needs to operate with $2^n\times 2^n$ matrices, i.e., exponentially many elements. The approach we present in this work is based on how the Pauli noise operators of the noise maps change or are affected by the operations of the manipulations. We update not only the pure graph state according to known, efficient rules, but also the noise operators. The goal is to be able to update the noise maps after each manipulation such that in the end there is a series of maps that act on the noiseless manipulated graph state. We refer to this approach as the noisy stabilizer formalism. This approach is equivalent to the straightforward one. Importantly, our approach allows for the analytical study of any kind of graph state and it poses an important advantage if the size of the final state is small. This is because the final updated noise maps are formed by Pauli noise operators of the size of the manipulated state. Moreover, there will be $2^m$ possible Pauli noise operators, where $m$ is the number of qubits of the manipulated state.

All the manipulation operations have been previously described in Sec.~\ref{ssec:graph:states}. In addition, we constantly make use of the fact that the action of all Pauli noise operators on a graph state can be equivalently described by products of Pauli $Z$ operators. Moreover, in this section $|G\rangle$ denotes an arbitrary $n$-qubit graph state and its density matrix $\rho=|G\rangle\langle G|$; $|G'\rangle$ denotes the resulting graph state after a sequence of manipulations of $m$ qubits and its density matrix is represented by $\rho'=|G'\rangle\langle G'|$.

\subsection{Commutation relations}
Consider a manipulation operator $O$, which can be any of the ones described in Sec.~\ref{ssec:graph:states}, and the Pauli noise operator $N$, which is of the shape $Z_j^{\alpha}Z_{N_j}^{\beta}$ with $\alpha, \beta = 0,1$ and $j\in V$. Note that the phase factor is disregarded as it is irrelevant for Pauli diagonal noise channels that we consider throughout this work. We apply the manipulation on the noisy state $ON|G\rangle$. Thus, we are interested in the commutation relation $ON=\tilde{N}O$ such that $ON|G\rangle = \tilde{N}O|G\rangle = \tilde{N}|G'\rangle$, where $\tilde{N}$ is in the Pauli group and denotes the updated noise operator that acts on the noiseless manipulated graph state, which can be written again with only Pauli $Z$ operators. Therefore, to update the Pauli noise operators of the noise maps, we compute the commutation relations between noise operators and manipulation operators.

\begin{table}[ht]
\centering
\begin{tabular}{cccl}
$\;$ Manipulation $O\;$                                     & \multicolumn{3}{c}{Noise $\tilde{N}$}             \\ \hline \hline
\multicolumn{1}{c|}{\multirow{3}{*}{$U_a^{\tau}$}}                                  & $Z_a^{\alpha}Z_{N_a}^{\alpha + \beta}$  & for & $j=a$       \\
\multicolumn{1}{c|}{}                                                               & $Z_b^{\alpha+\beta}Z_{N'_b}^{\beta}$          & for & $j=b\in N_a$    \\
\multicolumn{1}{c|}{}                                                               & $Z_j^{\alpha}Z_{N_j}^{\beta}$          & for & $j\neq a,b$    \\ \hline
\multicolumn{1}{c|}{\multirow{2}{*}{$L_{z,\pm}^{(a)}$}}                             & $Z_{N_a}^{\beta}$                       & for & $j=a$         \\
\multicolumn{1}{c|}{}                                                               & $Z_j^{\alpha}Z_{N'_j}^{\beta}$          & for & $j\neq a$      \\ \hline
\multicolumn{1}{c|}{\multirow{3}{*}{$L_{y,\pm}^{(a)}$}}                             & $Z_{N_a}^{\alpha + \beta}$              & for & $j=a$          \\
\multicolumn{1}{c|}{}                                                               & $Z_b^{\alpha+\beta}Z_{N'_b}^{\beta}$          & for & $j=b\in N_a$  \\
\multicolumn{1}{c|}{}                                                               & $Z_j^{\alpha}Z_{N'_j}^{\beta}$          & for & $j\neq a, b$      \\ \hline
\multicolumn{1}{c|}{\multirow{3}{*}{$L_{x,\pm}^{(a)}$}}                             & $\; Z_{b_0}^{\alpha}Z_{N_{b_0}}^{\alpha}$ & for & $j=a$        \\
\multicolumn{1}{c|}{}                                                               & $Z_{b_0}^{\beta}Z_{N'_{b_0}}^{\alpha}$   & for & $j = b_0$      \\
\multicolumn{1}{c|}{}                                                               & $Z_j^{\alpha}Z_{N'_j}^{\beta}$          & for & $j\neq a, b_0$ \\ \hline
\multicolumn{1}{c|}{\multirow{3}{*}{$L_{z,\pm}^{(t)}\text{CNOT}_{s\rightarrow t}$}} & $Z_s^{\alpha}Z_{N_t}^{\beta}$           & for & $j = t$        \\
\multicolumn{1}{c|}{}                                                               & $Z_j^{\alpha}Z_{N_j}^{\beta}$          & for & $j = s$ \\
\multicolumn{1}{c|}{}                                                               & $Z_j^{\alpha}Z_{N'_j}^{\beta}$          & for & $j \neq t,s$     \\ \hline \hline
\end{tabular}
\caption{Update rules for noise operators of the form $N= Z_j^{\alpha} Z_{N_j}^{\beta}$ with $\alpha, \beta = 0,1$. The left column specifies the manipulation operator $O$ (local complementation, Pauli measurements and merging), while the right column shows the updated noise operators $\tilde{N}$. This noise corresponds to the one after the manipulation $O$ on a graph state with initial noise operator $N\propto Z_j^{\alpha} Z_{N_j}^{\beta}$ with $\alpha, \beta = 0,1$, such that $\tilde{N}O=ON$. Here $N'_j$ ($N_j$) denotes the neighborhood of $j$ after (before) the manipulation. Recall that $Z_j^2=I_j$ and similarly $Z_{N_j}^2=I_{N_j}$. Update rules disregard phase factors, as they are irrelevant for Pauli diagonal noise channels that we consider throughout this work.}
\label{tab:commutations}
\end{table}

\begin{figure}[ht]
    \centering
    \includegraphics[width=\columnwidth]{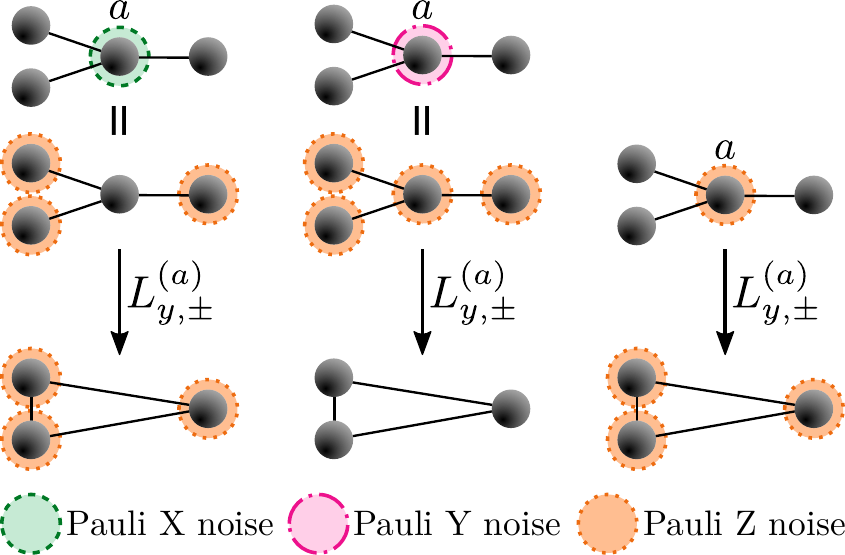}
    \caption{Graphical representation of a local Pauli measurement in the $Y$ basis on a qubit subject to Pauli noise $X$, $Y$ or $Z$.}
    \label{fig:manipulation:y}
\end{figure}

\begin{figure}[ht]
    \centering
    \includegraphics[width=\columnwidth]{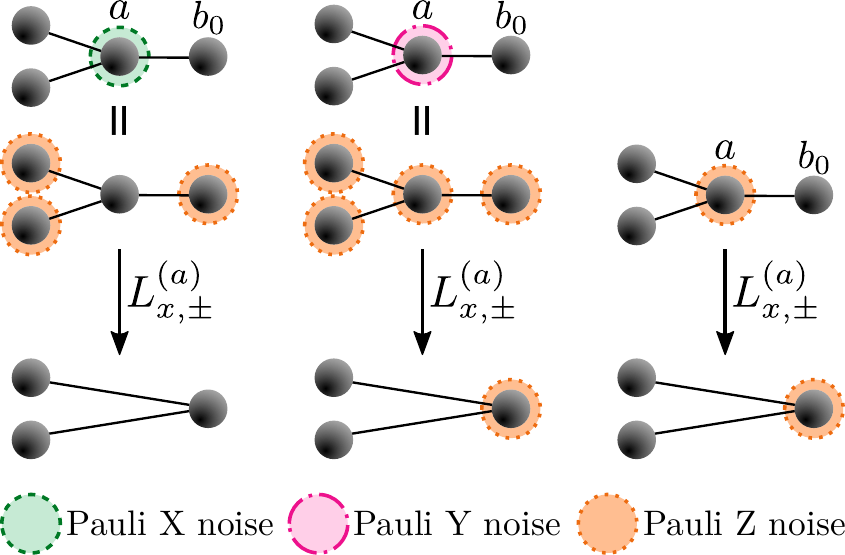}
    \caption{Graphical representation of a local Pauli measurement in the $X$ basis on a qubit subject to Pauli noise $X$, $Y$ or $Z$.}
    \label{fig:manipulation:x:a}
\end{figure}

In Table~\ref{tab:commutations}, the results of the update of noise operators after a certain manipulation are presented. The detailed computations of the results are presented in Appendix~\ref{a:nsf}, in which the examples for single-qubit Pauli noise and depolarizing noise are also included. Importantly, one can see from Table~\ref{tab:commutations}, that some noise operators of the qubits not directly involved in the manipulations change such that their structure is the same, but with the corresponding neighborhood after the manipulation. This is such that the overall noise pattern is not modified. However, the noise operators for the involved qubits have a more substantial change, such that the noise pattern is not maintained. This leads to the fact that sequential manipulations on neighboring qubits on a noisy graph state do not commute, meaning that the order in which certain manipulations are performed is relevant for the final overall noise pattern if they are performed in neighboring qubits. Note that this order relevance does not affect the Pauli $Z$ measurement. It is crucial to see that this property affects directly manipulation protocols widely used in entanglement-based quantum networks, e.g., creating bipartite entanglement from a large cluster \cite{raussendorf_one, vanmeter_recursive, dur_stability_2004, hahn_quantum_2019, matsuo_2018}, which requires sequential Pauli $Y$ or $X$ measurements of neighboring qubits. In Sec.~\ref{sec:application}, a simple example is presented to show this property and its importance explicitly. In Figs.~\ref{fig:manipulation:y} and \ref{fig:manipulation:x:a}, we graphically show how the different Pauli noise operators on a qubit change after local Pauli measurements in the $Y$ and $X$ bases on it, correspondingly. In Appendix~\ref{a:nsf}, we present the graphical representation of the update of noise operators under the remaining manipulations.

\subsection{Methodology}
Consider an $n$-qubit graph state $\rho=|G\rangle\langle G|$, where each qubit is subjected to single-qubit Pauli noise. Then this noisy graph state is manipulated by $k$ different operations, e.g., local complementations or Pauli measurements, where operations act on a set of qubits ($b_1, \dots, b_k$). Thus, the final state is given by
\begin{equation}
    O_k^{(b_k)}\cdots \, O_1^{(b_1)}\left(\mathcal{E}_n\cdots \mathcal{E}_1\rho\right){O_1^{\dagger}}^{(b_1)}\cdots \, {O_k^{\dagger}}^{(b_k)}.
\end{equation}
Using the noisy stabilizer formalism this can be equivalently described as 
\begin{equation}
    \widetilde{\mathcal{E}}_n\cdots \widetilde{\mathcal{E}}_1\left(O_k^{(b_k)}\cdots \, O_1^{(b_1)}\rho \, {O_1^{\dagger}}^{(b_1)}\cdots \, {O_k^{\dagger}}^{(b_k)}\right),
\end{equation}
where $\rho' = O_k^{(b_k)}\cdots \,O_1^{(b_1)}\rho \,{O_1^{\dagger}}^{(b_1)}\cdots\, {O_k^{\dagger}}^{(b_k)}$ is the noiseless manipulated state $\rho'=|G'\rangle\langle G'|$ and $\widetilde{\mathcal{E}}_i$ are the updated noise channels. Each of the noise channels is updated independently of each other as they all commute with each other when applied to graph states due to the fact that they can all be expressed as Eq.~\eqref{eq:pauli:map}. 

\begin{table}[ht]
\centering
\begin{tabular}{p{0.95\linewidth}}
\multicolumn{1}{c}{\textbf{Method:} For multiple manipulations} \\ \hline \hline
\hspace{1em}\textit{First manipulation:} Take each Pauli noise operators of each noise map
    $$N_i^{(j)}\propto Z_j^{\alpha(i)}Z_{N_j}^{\beta(i)},$$
where $j=1, \, \dots, \, n$ and $i=0,\,1,\,2,\,3$. Update them using the update rules $O_1N_i^{(j)}=\tilde{N}_i^{(j)}O_1$
    $$N_i^{(j)}\xrightarrow{O_1^{(b_1)}} \tilde{N}_i^{(j)}$$
The updated noise operators are a product of Pauli $Z$ operators.
\\
\hspace{1em}\textit{Second manipulation:} Update each Pauli noise operators of each noise map using the update rules
    $$\tilde{N}_i^{(j)}\xrightarrow{O_2^{(b_2)}} \tilde{\tilde{N}}_i^{(j)}$$
where $j=1, \, \dots, \, n$ and $i=0,\,1,\,2,\,3$. Once again, the updated noise operators are a product of Pauli $Z$ operators.
    $$\vdots$$
\hspace{1em}\textit{Last manipulation $O_k$:} Update each Pauli noise operators of each noise map using the update rules
    $$\tilde{\stackrel{.}{\stackrel{.}{\stackrel{.}{\tilde{\tilde{N}}}}}}_i^{(j)} \xrightarrow{O_k^{(b_k)}} \widetilde{N}_i^{(j)} := \tilde{\tilde{\stackrel{.}{\stackrel{.}{\stackrel{.}{\tilde{\tilde{N}}}}}}}_i^{(j)}  $$
where $j=1, \, \dots, \, n$ and $i=0,\,1,\,2,\,3$. Once again, the updated noise operators are a product of Pauli $Z$ operators. 
\\ \hline \hline
\end{tabular}

\caption{Step-by-step guide on how to use the noisy stabilizer formalism to update the Pauli noise operators under the action of a series of manipulations. The mentioned update rules can be found in Table~\ref{tab:commutations}. }
\label{tab:method}
\end{table}
In Table~\ref{tab:method} we present a guided procedure on how to make use of the noisy stabilizer formalism for a set of manipulations on the same graph state. Note that the graph state changes after each manipulation and so do the Pauli noise operators, so each manipulation depends on the previous one to update both the graph state and the Pauli noise operators. Moreover, at each manipulation, we update each Pauli noise operator from each map independently, which is allowed because all Pauli noise operators and noise maps commute with each other. After applying the procedure described in Table~\ref{tab:method}, the Pauli noise operators have been updated $k$ times, such that 
\begin{equation}
    N_i^{(j)}\xrightarrow{O_1^{(b_1)} \cdots \,O_k^{(b_k)}} \widetilde{N}_i^{(j)}
\end{equation}
where $j=1, \, \dots, \, n$ and $i=0,\, 1,\, 2,\, 3$. Thus, all the noise maps can be finally updated using the updated noise operators, such that
\begin{equation}
    \mathcal{M}_j\rho \xrightarrow{O_1^{(b_1)}\cdots \, O_k^{(b_k)}} \widetilde{\mathcal{M}}_j\rho' 
\end{equation}
for $j=1, \, \dots, \, n$, where $\mathcal{M}_j\rho$ corresponds to Eq.~\eqref{eq:pauli:map} and 
\begin{equation}
    \widetilde{\mathcal{M}}_j\rho'=\sum_{i=0}^3 \lambda_i \widetilde{N}_i^{(j)}\rho' {\widetilde{N}_i^{(j)}}{}^{\dagger}.
\end{equation}
Finally, one can apply all the final noise maps $\{\widetilde{\mathcal{M}}_j\}$ to the final noiseless graph state $\rho'$ to retrieve the final noisy graph state. Hence, the size of the final graph state directly impacts the size of the final Pauli noise operators as they should be of the same size. Furthermore, in \cite{dur_stability_2004, hein_entanglement_2005} it was stated that noise maps $\mathcal{M}_a$ only act nontrivially on $a$ and its neighbors. A manipulation can divide the initial graph states into several graphs. In \cite{dur_stability_2004, hein_entanglement_2005} it is mentioned that the noise of these resulting states can be obtained by considering only the (reduced) action of the noise maps that act nontrivially on the reduced graph state. Therefore, one does not need to update all the noise operators in the system, just the ones that act nontrivially.

\section{Generalization of the method}\label{sec:limits}
In Sec.~\ref{sec:noisy:stabilizer:formalism}, we presented an efficient formalism to describe single-qubit Pauli noise on graph states that are manipulated by Pauli measurements or Clifford operations. Now, we explore generalizations of the method and discuss limitations by studying its use on more general quantum systems and noise models.

\subsection{General noise models}\label{ssec:general:noise}
Consider a general noise model described by a completely positive map (CPM) $\mathcal{E}$ that acts on $n$ qubits of a graph state corresponding to a graph $G$, $\mathcal{E}|G\rangle\langle G|$.
Consider that this general channel can be decomposed in $\mathcal{E}_l\cdots \mathcal{E}_1$, where each CPM acts on a subset of qubits, which in the simplest case would be single-qubit channels. Whenever these noise maps are of multiqubit Pauli diagonal form, we can generalize our noisy stabilizer formalism to update the corresponding maps and provide an efficient description. Note that the range of the maps, i.e., the systems on which they act, do not need to be distinct; qubits might be affected by multiple maps. 

\begin{figure}[ht]
    \centering
    \includegraphics[width=0.9\columnwidth]{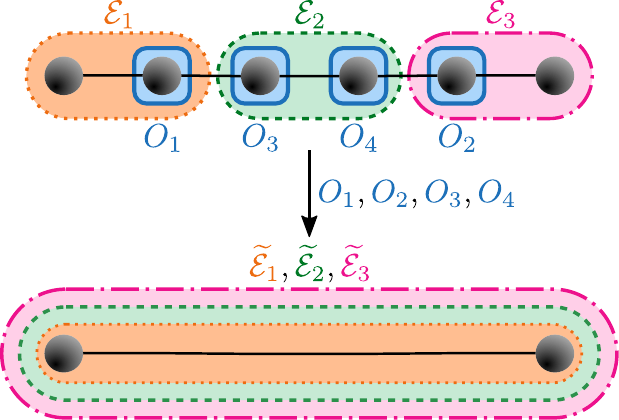}
    \caption{Graphical representation of the use of the noisy stabilizer formalism for general diagonal Pauli noise modes. The top represents a graph state subject to several Pauli diagonal noise maps denoted by $\mathcal{E}_i$, which is then manipulated by operators $O_j$ on the corresponding qubits [as described in Eq.~\eqref{eq:before:nsf}]. If the manipulations are performed it is equivalent to having the noiseless manipulated state subject to the updated noise maps denoted by $\widetilde{\mathcal{E}}_i$ [as described in Eq.~\eqref{eq:after:nsf}].}
    \label{fig:general:noise}
\end{figure}

We consider now the situation where the initial noisy graph state is manipulated by $k$ different operations, where operations act on a set of qubits $(b_1, \dots, b_k)$. Thus, the final state is described by
\begin{equation}\label{eq:before:nsf}
     O_k^{(b_k)}\cdots \, O_1^{(b_1)}\left(\mathcal{E}_l\cdots \mathcal{E}_1\rho\right){O_1^{\dagger}}^{(b_1)}\cdots \, {O_k^{\dagger}}^{(b_k)}.
\end{equation}
Now, each of the noise channels $\mathcal{E}_k$ can be updated using the noisy stabilizer formalism independently. For this, it is crucial that maps are Pauli diagonal. Therefore, the final state can be also expressed as
\begin{equation}\label{eq:after:nsf}
    \widetilde{\mathcal{E}}_l\cdots \widetilde{\mathcal{E}}_1\left(O_k^{(b_k)}\cdots \, O_1^{(b_1)}\rho \, {O_1^{\dagger}}^{(b_1)}\cdots \, {O_k^{\dagger}}^{(b_k)}\right),
\end{equation}
where $\rho' = O_k^{(b_k)}\cdots \,O_1^{(b_1)}\rho \,{O_1^{\dagger}}^{(b_1)}\cdots\, {O_k^{\dagger}}^{(b_k)}$ is the noiseless manipulated state, $\rho'=|G'\rangle\langle G'|$, and $\widetilde{\mathcal{E}}_i$ are the updated noise channels. This transformation of the noise maps is also described graphically in Fig.~\ref{fig:general:noise}. The overall effort and computational complexity are determined by the number of terms in the noise channels $\mathcal{E}_k$ when written in the Pauli basis. If the noise channels have few terms each, the update is efficient. This is the case whenever each noise channel is diagonal in the Pauli basis and acts on a few qubits only or if only a few terms appear as is, e.g., the case for correlated noise, as detailed below.

\subsubsection{Correlated noise models}
A correlated Pauli noise channel can be expressed as
\begin{equation}\label{eq:correlated:pauli}
    \mathcal{E}\rho =p\rho+(1-p)\sigma_i^{\otimes k} \rho \,\sigma_i^{\otimes k},
\end{equation}
where $\sigma_i^{\otimes k}$ acts on all the qubits in the system or a particular subset of them. Note that these noise operators can also be expressed in terms of only Pauli $Z$ operators when they are applied to graph states. Therefore, the commutation relations presented in Table~\ref{tab:commutations} can be used to update these correlated noise operators. In particular, one has to compute the update of $\sigma_i$ on each qubit that the correlated operator acts on; thus a total of $k$ updates have to be computed for Eq.~\eqref{eq:correlated:pauli}. These statements are also true for any correlated Pauli noise channel. However, this formalism is not suited to analyze true correlated noise models such that $\mathcal{E}\rho = \int d\delta t \, e^{-iH\delta t}\rho \,e^{iH\delta t}$, where $H=\sum_i Z_i$.

In \cite{buterakos_deterministic} a method to deterministically generate an arbitrary photonic graph state was proposed. From this one can see that the noise arising from the graph state generation would be a correlated noise model with noise operators that do not act always on the entire graph state. Nonetheless, it is important to remember that one only needs to update the noise operators that act nontrivially on the qubits that will be manipulated and their neighborhoods, which might decrease the number of noise operators to update. In general, the overall effort is the same as for the noncorrelated noise models, as the bottleneck is determined by the size of the manipulated graph state.

\subsection{Stabilizer states}
The use of this formalism can be further expanded to any stabilizer state. This is due to the fact that a stabilizer state is equivalent to a graph state under local Clifford operations  \cite{nest_graphical}. Consider Pauli noise acting on a stabilizer state. Next, the local Clifford operation to transform it to a graph state is applied to the noisy stabilizer state. Then, if the commutation between the Pauli noise operators and the local Clifford operation is computed, one has a new set of Pauli noise operators that act on a graph state that corresponds to a certain stabilizer state \cite{nest_invariants_2005}. These new noise operators can be studied and treated as seen in the presented noisy stabilizer formalism, as they are Pauli noise operators acting on a graph state.

In this way, also manipulation by Clifford operations and Pauli measurements on general stabilizer states can be taken into account, thereby extending the stabilizer formalism and the Gottesman-Knill theorem also to noisy scenarios. 

\section{Efficiency of the method}\label{sec:efficiency}
As one can see in Table~\ref{tab:method}, in the most general case we need to compute the total update of each Pauli noise operator per each qubit, that is, $4n$ total updates for an $n$-qubit graph state. However, for some particular noise models, the update of not all Pauli noise operators has to be computed. The correlated noise model presented in Eq.~\eqref{eq:correlated:pauli} is an example of this, as all qubits involved are subject to the same kind of Pauli noise. 

Besides computing a maximum of $4n$ updates, one needs to consider that an effort has to be made for each term for all noise maps acting on the system. This effort is not in terms of computation difficulty but in terms of memory. Therefore, the total number of terms of all maps establishes an overhead for the use of this formalism. 

Nevertheless, the main bottleneck to the use of this formalism is the size of the final manipulated graph states, since the final updated noise operators for each noise map are of the size of the final graph state. Moreover, if all maps are applied, it results in a maximum of $2^{2m}$ possible terms acting on the final $m$-qubit graph state. 

\section{Application of the method}\label{sec:application}
In this section, we illustrate the application of the method and show explicitly the advantages and results of the presented noisy stabilizer formalism via a simple example. As the initial graph state, we take an $N$-qubit 1D cluster, where each qubit is labeled by a number from 1 to $N$. The target state we want to achieve is a two-qubit graph state between qubits 1 and $N$. The manipulation that needs to be done is the Pauli $Y$ measurement of all qubits but 1 and $N$.

We consider that all the qubits in the initial graph state are subject to depolarizing noise. Therefore, the noise maps can be computed following Eq.~\eqref{eq:pauli:map} and are of the form
\begin{equation}\label{eq:depolar:map}
    \mathcal{M}_a\rho=p\rho+\frac{1-p}{4}\sum_{\alpha, \beta =0}^{1}\left(Z_a^{\alpha}Z_{N_a}^{\beta}\right)\rho \left(Z_a^{\alpha}Z_{N_a}^{\beta}\right).
\end{equation}
Using the updating rules of Table~\ref{tab:commutations}, one can see that the final noise maps of the target pair qubits are described as 
\begin{equation}\label{eq:map:target}
    \mathcal{M}_{\text{target}}\rho' = p\rho' + \frac{1-p}{4}\sum_{\alpha, \beta =0}^{1}Z_1^{\alpha}Z_{N}^{\beta}\rho' Z_1^{\alpha}Z_{N}^{\beta}.
\end{equation}
Moreover, the noise maps of the qubits that have been measured in a Pauli basis can take one of the forms
\begin{align}
    \mathcal{M}_{\alpha}\rho'& = p\rho' + \frac{1-p}{2}\left(\rho' + Z_1 Z_N\rho' Z_1 Z_N\right), \label{eq:ma} \\
    \mathcal{M}_{\beta}\rho'& = p\rho' + \frac{1-p}{2}\left(\rho' + Z_1\rho' Z_1\right), \label{eq:mb}\\
    \mathcal{M}_{\gamma}\rho'& = p\rho' + \frac{1-p}{2}\left(\rho' + Z_N\rho' Z_N\right). \label{eq:mc}
\end{align}
It is convenient to define a vector $\boldsymbol{w}=(w_{\alpha}, w_{\beta}, w_{\gamma})$, where $w_i$ denotes the weight of the final noise map $\mathcal{M}_i$ with $i=\alpha, \beta, \gamma$. We refer to this vector $\boldsymbol{w}$ as the \textit{weight vector}. As shown in Sec.~\ref{sec:noisy:stabilizer:formalism}, the order of consecutive Pauli $Y$ measurements in a noisy graph state is relevant for the final noise pattern, and thus different measurement patterns will lead to different weight vectors. However, the qubits that are fully merged do not have this shape; instead, their noise maps have the same form as the noise maps of the target qubits, defined in Eq.~\eqref{eq:map:target}. We define the parameter $t$ as the number of full merging operations, which involve two qubits each, in a certain manipulation.

\begin{figure}\label{fig:example}
    \centering
    \subfloat[]{\includegraphics[width=0.46\columnwidth]{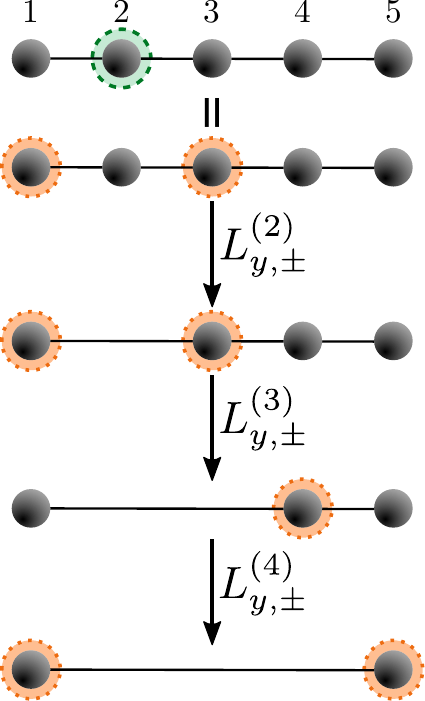}\label{fig:example:x}}
    \hspace{0.19in}
    \subfloat[]{\includegraphics[width=0.46\columnwidth]{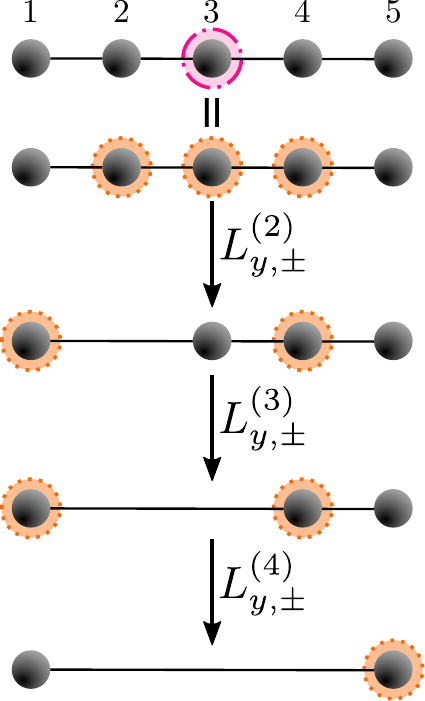}\label{fig:example:y}}
    \hfill
    \subfloat{\includegraphics[width=0.9\columnwidth]{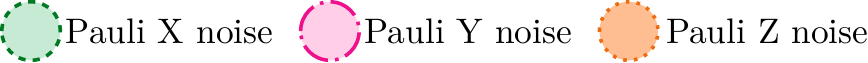}}
    \caption{Graphical representation of a five-qubit 1D cluster, where qubits 2, 3 and 4 are measured sequentially in the $Y$ basis. The evolution or updating of a certain noise operator in the cluster is represented during the considered measurements: (a) qubit 2 is subject to Pauli $X$ noise and in (b) qubit 3 is subject to Pauli $Y$ noise.}
\end{figure}

Note that the final noiseless state is a Bell state in the graph state basis and that the action of the nonidentity Kraus operators changes it to another Bell state (in the graph basis). Therefore, only the part where the Kraus operators are identities will have a nonzero contribution to the total fidelity. Thus, one can compute the fidelity given a certain weight vector $\boldsymbol{w}$ and the number of full merges $t$ and is
\begin{equation}\label{eq:general:fidelity}
    F(p,\boldsymbol{w}, t)=\frac{1}{4}\left(1+p^{2+2t}\sum_{\substack{i,j=\alpha, \beta, \gamma \\ i\neq j}}p^{w_i + w_j}\right).
\end{equation}
Moreover, from Eq.~\eqref{eq:general:fidelity} one can see that the contribution of each noise map of the measured qubits to the fidelity is the same, meaning that the order of the components in the weight vector does not matter. Note that in this example, only local Pauli measurements in the $Y$ basis are performed; thus $t=0$. In Fig.~\ref{fig:example} we show the evolution of a certain noise operator on an instance with a five-qubit 1D cluster. In particular, in Fig.~\ref{fig:example:x} we consider Pauli $X$ noise on the second qubit and we show how it evolves under each local Pauli $Y$ measurement. Similarly, in Fig.~\ref{fig:example:y} we consider Pauli $Y$ noise on the third qubit. Note that Fig.~\ref{fig:example} does not include all the noise operators and their evolution considered in this example.

Note that if one considers any other initial graph state that is manipulated into a Bell pair, as in this example, the (reduced) noise maps of the manipulated qubits that act on the (reduced) Bell pair take the forms described above. Thus, the description using $\boldsymbol{w}$ and $t$ and the fidelity described in Eq.~\eqref{eq:general:fidelity} are valid as long as the final (reduced) graph state is a Bell pair. 
 
\subsection{Measuring strategies}\label{ssec:strategies}
We have analyzed three different strategies to perform the Pauli $Y$ measurements in the initial 1D cluster. 

\subsubsection{Strategy - side to side}
This strategy consists in measuring the qubits consecutively, one by one, from one end of the cluster to the other. The weight vector for this strategy is
\begin{equation}\label{eq:side:to:side}
    \boldsymbol{w}=\left(\frac{n+g(n)}{2}, 0,\frac{n-g(n)}{2}\right),
\end{equation}
where $n=N-2$ is the number of measured qubits and $g(n) = [1-(-1)^n]/2$. The proof of this expression can be found in Appendix~\ref{a:side:to:side}.

\subsubsection{Strategy - every second qubit}
This strategy allows for simultaneous measurements as one would first measure all even (or odd) qubits, which is allowed as they are nonconsecutive qubits. Then one would keep measuring every second qubit until all measurements are performed. The weight vector for this strategy is 
\begin{equation}\label{eq:every:second:qubit}
    \boldsymbol{w}= \begin{cases}
    (l,\, l,\, l) & \text{ for } n=3l \\
    (l,\, l,\, l+1) & \text{ for } n=3l+1 \\
    (l,\, l+1,\, l+1) & \text{ for } n=3l+2
    \end{cases}
\end{equation}
where $l\in \mathbb{Z}^+$ and $n=N-2$. This weight vector can only be used for $n>2$. However, if $n=0,1,2$ one can use Eq.~\eqref{eq:side:to:side}, as the two strategies are the same for these values of $n$.

\subsubsection{Strategy - pairs} 
The last strategy consists in measuring two qubits in each step from the outside to the inside, only if they are nonconsecutive. First, qubits 2 and $N-1$ are measured, then 3 and $N-2$, and so on. The corresponding weight vector is 
\begin{equation}\label{eq:pairs}
    \boldsymbol{w}= \begin{cases}
    (l,\, l,\, 2l-1) & \text{ for } n=4l-1 \\
    (l,\, l,\, 2l) & \text{ for } n=4l \\
    (l,\, l,\, 2l+1) & \text{ for } n=4l+1 \\
    (l,\, l +1,\, 2l+1) & \text{ for } n=4l+2 \\
    \end{cases}
\end{equation}
where $l\in \mathbb{Z}^+$ and $n=N-2$. This weight vector can only be used for $n>2$. However, if $n=0,1,2$ one can use Eq.~\eqref{eq:side:to:side}, as the two strategies are the same for these values of $n$.

\subsection{Comparison between strategies}
\begin{figure}[h]
    \centering
    \includegraphics[width=\columnwidth]{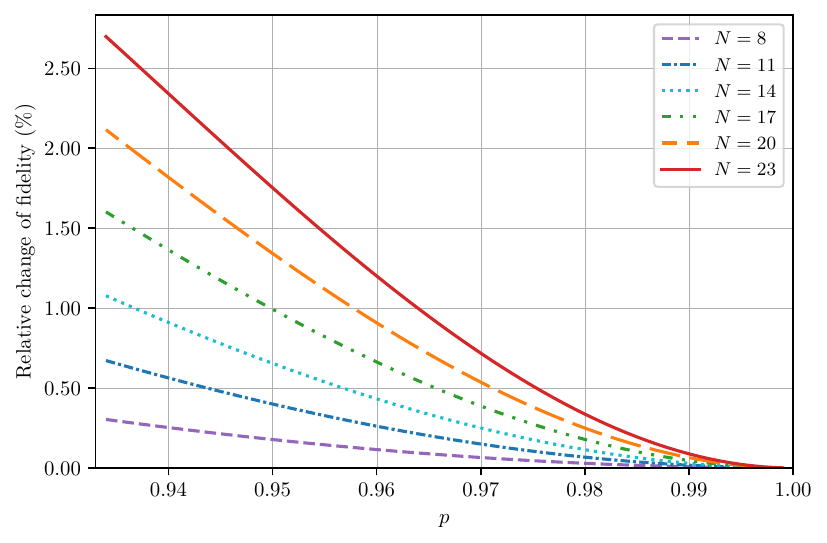}
    \caption{Linear plot of the relative change of the fidelity in percentage between two strategies (every-second-qubit and side-to-side) to achieve a two-qubit graph state obtained from an $N$-qubit 1D cluster subject to depolarizing noise with probability $1-p$. The $y$ axis corresponds to $\Delta F / F_{\text{esq}}$, where $F_{\text{esq}}$ is the fidelity of the every-second-qubit strategy and $\Delta F$ corresponds to the difference between the fidelity for the side-to-side strategy and the every-second-qubit strategy. Note that the latter corresponds to the best and the worst strategies in terms of fidelity. Only the regime where $F_{\text{esq}}$ is above 0.5 is plotted, restricting the regime of the $x$ axis. Each line corresponds to a different size of the initial 1D cluster.}
    \label{fig:relative:change:fidelities}
\end{figure}

Using the weight vectors presented above and the expression for the fidelity in Eq.~\eqref{eq:general:fidelity}, we compute the fidelity for each strategy in Appendix~\ref{a:fidelity}. For some values of $n$, the every-second-qubit and pairs strategies have the same fidelity. Nevertheless, in general, the side-to-side strategy has a higher fidelity in terms of $p$, whereas the every-second-qubit strategy has the worst one. However, the differences between them are small. To see this, we show in Fig.~\ref{fig:relative:change:fidelities} the relative change of the fidelity in percentage between the side-to-side and every-second-qubit strategies for a fixed size of the initial cluster $N=n+2$. Notice that for larger $N$ the relative change increases. Moreover, the relative change is more relevant as the probability of noise increases. Nonetheless, for values of fidelity slightly above 0.5, the relative change goes from roughly 0.3\% to 2.7\%.

\section{Conclusions and outlook}\label{sec:conclusion}
In this paper we have generalized the stabilizer formalism for pure states to a noisy one, which can also treat the action of Pauli diagonal noise processes on stabilizer states efficiently. The key observation is that for stabilizer or graph states, the action of Pauli noise operators on such states can be efficiently described and updated whenever Clifford operations or local Pauli measurements are performed. This implies that for any such process, the influence of (quasi) local noise on the resulting state can be determined efficiently. 

Given the fact that many relevant protocols and methods in quantum information processing are based on graph states, Clifford operations and Pauli measurements, and that local Pauli noise is a generic and relevant error model to describe the influence of noise, imperfections and decoherence on such systems, this provides a widely applicable and powerful method to study noisy protocols. Protocols based on graph states and their manipulation by Clifford operations include quantum error correction schemes \cite{nielsen_chuang_2010, stabilizer_gottesman}, entanglement purification \cite{bennet_mixed, bennet_purification, deutsch_quantum, dur_entanglement_2003} and the transformation of multipartite entangled resource states to some target states. Similarly, also measurement-based quantum computation of Clifford circuits \cite{briegel_measurement_based_2009, raussendorf_measurement} falls into this category. While we have only illustrated our method by a simple example of generating a Bell state from a 1D cluster state, we believe that the formalism we developed can also be applied to the analysis of the influence of noise in entanglement-based quantum networks \cite{pirker_modular_2018, pirker_quantum_2019, pirker_construction_2017, miguel-ramiro_optimized_2021}, which we will report on a separate work \cite{mor_influence}. In the latter, this formalism is directly employed to obtain the fidelity of a target state, e.g., a Bell pair or a three-qubit Greenberger-Horne-Zeilinger state, achieved from a certain resource state that ranges from a large multidimensional cluster to a collection of smaller states, for instance, Bell pairs. Further applications in the context of error analysis for fault-tolerant computation, quantum error correction codes, or bipartite and multipartite entanglement purification protocols acting on multiple copies are conceivable. 

Nonetheless, the method is limited to graph states and diagonal noise processes in the Pauli basis. The treatment of generic noise processes seems to involve further difficulties, as phases for nondiagonal Pauli noise terms do not cancel, and independent treatment of noise maps seems to be impossible. Furthermore, to explicitly evaluate the resulting state and, e.g., determine state or process fidelities, it seems that at some point one needs to switch to an explicit description and actually compute the action of noise maps on the state, i.e., work with density matrices, which involves an effort that scales exponentially in the size of the final state. This is true even though the description of the states and noise maps themselves is efficient, as they scale linearly in the total number of systems involved. Nevertheless, as long as target states are small, i.e., consist only of a few qubits, the formalism provides a powerful tool to describe the influence of noise and study noisy quantum information processing protocols.  

\section*{Acknowledgements}
This work was supported by the Austrian Science Fund (FWF) through Projects No. P30937-N27, No. P36009-N and No. P36010-N and Finanziert von der Europäischen Union - NextGenerationEU.

\bibliographystyle{apsrev4-1}
\bibliography{refs_PRA_v3.bib}
\clearpage
\renewcommand\appendixname{Appendix}
\appendix
\onecolumngrid
\section{Update rules for noise operators}\label{a:nsf}
In this appendix we provide the detailed computations of the commutations presented in Table~\ref{tab:commutations}. Moreover, we also show the noise maps for the case of single-qubit depolarizing noise on all qubits in a graph state.

\subsection{Local complementation}\label{a:nsf:local:complementation}
In Eq.~\eqref{eq:local:complementation}, the graph state after the local complementation on $a$ is presented. Note that for the noise operator $Z_j$ when $j\neq a$, $[Z_j, U_{a}^{\tau}]=0$. For $j=a$, the noise operator $Z_a$ does not commute with the manipulation operator. To understand how the noise operator impacts the manipulation, we compute the commutation between $U_{a}^{\tau}$ and $Z_a$. Then $U_{a}^{\tau}Z_a|G\rangle = -Y_a U_{a}^{\tau}|G\rangle$, so $Y_a$ is acting on the manipulated graph state. Remember that $Y_a$ is the same as $Z$ on $a$ and its neighbors with a phase $-i$. Recall that the effect of local complementation on $a$ alters the neighborhood of the neighbors of $a$, but not the neighborhood of $a$. Thus, 
\begin{equation}
    U_{a}^{\tau}Z_a|G\rangle = i Z_a Z_{N_a}|G'\rangle.
\end{equation}
Therefore, when there is noise on the qubit where the local complementation is applied, this noise is spread to its neighbors.

With only these commutation relations of $Z$ noise on any qubit on the state, we can know how any noise operator of a noise map is transformed since we can express the action of a map by only Pauli $Z$ operators. 
\begin{table}[ht]
\centering
\begin{tabular}{|clc|}
\hline
Elements of $\mathcal{M}_a \rho$ &               & Elements of $\widetilde{\mathcal{M}}_a \rho'$ \\ \hline \hline
$\rho $         & $\rightarrow$ &  $\rho'$           \\
$X_a\rho X_a = Z_{N_a} \rho Z_{N_a}$         & $\rightarrow$ &  $Z_{N_a} \rho' Z_{N_a}$           \\
$Y_a\rho Y_a = Z_a Z_{N_a} \rho Z_a Z_{N_a}$         & $\rightarrow$ & $Z_a \rho' Z_a$           \\
$Z_a \rho Z_a$         & $\rightarrow$ & $Z_a Z_{N_a} \rho' Z_a Z_{N_a}$           \\ \hline
\end{tabular}
\caption{Elements of the noise map of $a$ before, $\mathcal{M}_a\rho$, and after, $\widetilde{\mathcal{M}}_a\rho'=U_{a}^{\tau}(\mathcal{M}_{a}\rho)(U_{a}^{\tau})^{\dagger}$, a local complementation on $a$.}
\label{tab:local:complementation}
\end{table}

\begin{figure}[ht]
    \centering
    \includegraphics[width=0.5\columnwidth]{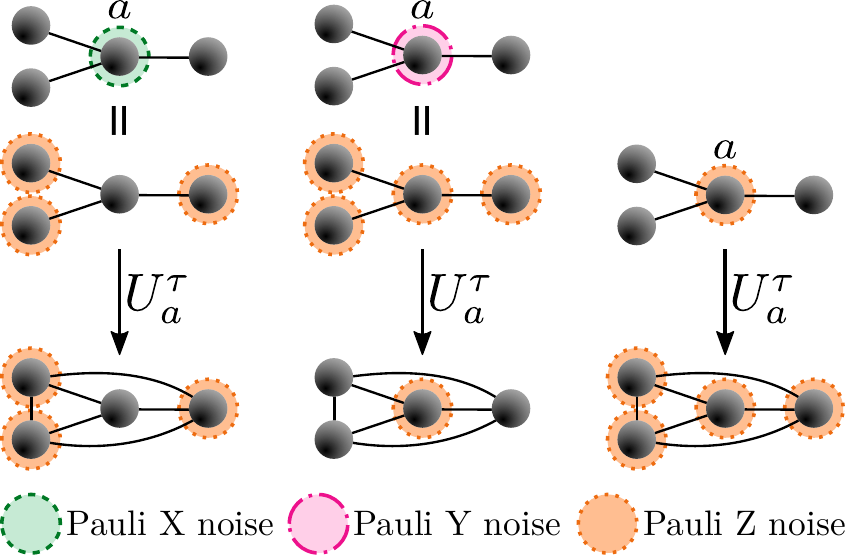}
    \caption{Graphical representation of a local complementation on a qubit subject to Pauli noise, $X$, $Y$ or $Z$.}
    \label{fig:manipulation:lc}
\end{figure}

Table~\ref{tab:local:complementation} shows the transformation of the elements of the noise map of the qubit where the local complementation is applied. Moreover, in Fig.~\ref{fig:manipulation:lc}, we graphically show how the different Pauli noise operators on a qubit change after local complementation. 

For any qubit $b\in N_a$, the elements of their noise maps change as
\begin{equation}\label{eq:kraus:in:Na}
    Z_b^{\alpha}Z_{N_b}^{\beta} \rho Z_b^{\alpha}Z_{N_b}^{\beta} \rightarrow Z_b^{\alpha+\beta}Z_{N'_b}^{\beta} \rho' Z_b^{\alpha+\beta}Z_{N'_b}^{\beta},
\end{equation}
where $\alpha, \beta = 0,1$ and $N'_j$ denotes the neighborhood of $j$ after the manipulation. For any qubit $j\neq a$ and $j\notin N_a$, the elements of their noise maps do not change and thus
\begin{equation}\label{eq:kraus:no:change}
    Z_j^{\alpha}Z_{N_j}^{\beta} \rho Z_j^{\alpha}Z_{N_j}^{\beta} \rightarrow Z_j^{\alpha}Z_{N_j}^{\beta} \rho' Z_j^{\alpha}Z_{N_j}^{\beta}
\end{equation}
where $\alpha, \beta = 0,1$ and $N_j$ denotes the neighborhood before the manipulation, as the neighborhood of qubit $j$ does not change after a local complementation on $a$, $N'_j = N_j$.

Notably, if depolarizing noise is considered, the noise maps of the graph state are transformed such that $U_{a}^{\tau}(\mathcal{M}_{j}\rho)(U_{a}^{\tau})^{\dagger}=\mathcal{M}_{j}\rho'$ for all $j\in V$. In other words, they are the same as if we would write them after the manipulation. This is because all the nonidentity Kraus operators have the same weight, and the depolarizing map can be equivalently written in any (updated) basis. 

\subsection{Local Pauli measurements} \label{a:nsf:local:pauli:measurements}
As seen in Sec.~\ref{ssec:graph:states}, local Pauli measurements are described by Eqs.~\eqref{eq:correction:z}-\eqref{eq:graph:x}. Importantly, from \cite{hein_multiparty_2004} we make use of the commutation relations
\begin{align}
    P_{x,\pm}^{(a)}Z_a & = Z_a P_{x,\mp}^{(a)}, \\
    P_{y,\pm}^{(a)}Z_a & = Z_a P_{y,\mp}^{(a)}, \\
    P_{z,\pm}^{(a)}Z_a & = Z_a P_{z,\pm}^{(a)}.
\end{align}
Additionally, the noiseless graph state after a local Pauli measurement is 
\begin{equation}
    |G'\rangle = L_{i,\pm}^{(a)}|G\rangle := \langle{i,\pm}|^{(a)}\otimes (U_{i,\pm}^{(a)})^{\dagger}P_{i,\pm}^{(a)}|G\rangle,
\end{equation}
where $i=x,y,z$ denotes the basis of the measurement and $L_{i,\pm}^{(a)}$ is described in Eq.~\eqref{eq:measured:graph}.

\subsubsection{Pauli Z measurement}\label{a:nsf:z:measurement}
\begin{figure}[ht]
    \centering
    \includegraphics[width=0.5\columnwidth]{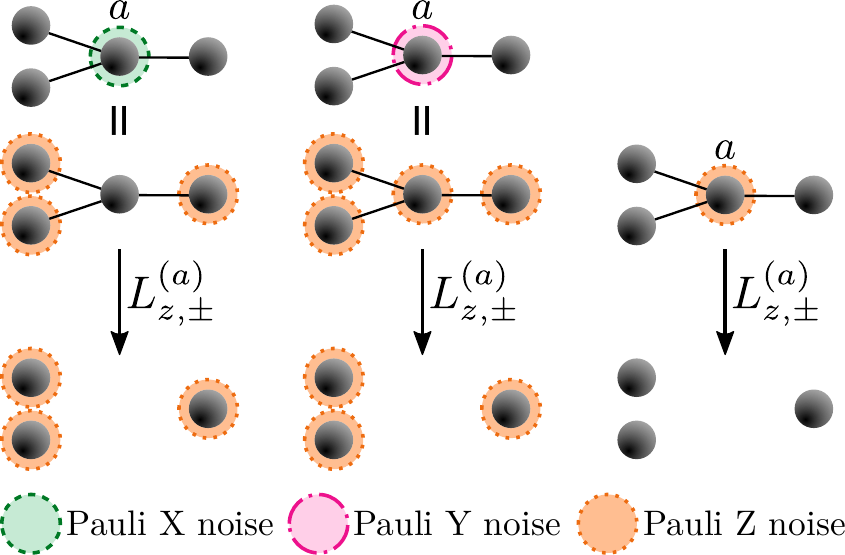}
    \caption{Graphical representation of a local Pauli measurement in the $Z$ basis on a qubit subject to Pauli noise, $X$, $Y$ or $Z$.}
    \label{fig:manipulation:z}
\end{figure}

The noise operator $Z_j$ commutes with $L_{z, \pm}^{(a)}$ for $j\neq a$. However, the noise operator $Z_a$ commutes with $(U_{z,\pm}^{(a)})^{\dagger}P_{z,\pm}^{(a)}$ and acts on $\langle{z,\pm}|^{(a)}$ by adding a general phase to the manipulated state depending on the measurement outcome, such that 
\begin{equation}\label{eq:pauli:z:a}
    L_{z, \pm}^{(a)} Z_a = \pm L_{z, \pm}^{(a)}.
\end{equation}
From Eq.~\eqref{eq:pauli:z:a} we get the results of Table~\ref{tab:pauli:z}, which show the transformations of the elements of the noise map of the measured qubit $a$. Moreover, in Fig.~\ref{fig:manipulation:z}, we graphically show how the different Pauli noise operators on a qubit change after the local Pauli $Z$ measurement on it.

\begin{table}[ht]
\centering
\begin{tabular}{|clc|}
\hline
Elements of $\mathcal{M}_a \rho$ &               & Elements of $\widetilde{\mathcal{M}}_a \rho'$ \\ \hline \hline
$\rho $         & $\rightarrow$ &  $\rho'$           \\
$X_a\rho X_a = Z_{N_a} \rho Z_{N_a}$         & $\rightarrow$ &  $Z_{N_a} \rho' Z_{N_a}$           \\
$Y_a\rho Y_a = Z_a Z_{N_a} \rho Z_a Z_{N_a}$         & $\rightarrow$ & $Z_{N_a} \rho' Z_{N_a}$           \\
$Z_a \rho Z_a$         & $\rightarrow$ & $\rho'$           \\ \hline
\end{tabular}
\caption{Elements of the noise map of $a$ before, $\mathcal{M}_a\rho$, and after, $\widetilde{\mathcal{M}}_a\rho'=L_{z,\pm}^{(a)}(\mathcal{M}_{a}\rho)(L_{z,\pm}^{(a)})^{\dagger}$, a local Pauli $Z$ measurement on $a$.}
\label{tab:pauli:z}
\end{table}

For any other qubit $j\neq a$, the elements of their noise maps change as 
\begin{equation}\label{eq:kraus:not:a}
    Z_j^{\alpha}Z_{N_j}^{\beta} \rho Z_j^{\alpha}Z_{N_j}^{\beta} \rightarrow Z_j^{\alpha}Z_{N'_j}^{\beta} \rho' Z_j^{\alpha}Z_{N'_j}^{\beta}
\end{equation}
where $\alpha, \beta = 0,1$. Note that the noise operators now act on the graph state corresponding to $G'$ after the measurement and $N'_j$ denotes the neighborhood after the $Z$ measurement. 

Notably, the depolarizing maps for $j \neq a$ will be the same as if we would write them after the measurement. However, the depolarizing noise map for $a$ results in
\begin{equation}\label{eq:depolar:map:a}
    \widetilde{\mathcal{M}}_a\rho'=p\rho'+\frac{1-p}{2}\left(\rho'+Z_{N_a}\rho'Z_{N_a}\right).
\end{equation}

\subsubsection{Pauli Y measurement}\label{a:nsf:y:measurement}
The noise operator $Z_j$ commutes with $L_{y, \pm}^{(a)}$ for $j\neq a$. However, the noise operator $Z_a$ does not, as
\begin{equation}\label{eq:pauli:y:a}
    L_{y, \pm}^{(a)} Z_a = \prod_{b\in N_a}(\pm i Z_b) L_{y, \pm}^{(a)}.
\end{equation}
Therefore, applying a local Pauli $Y$ measurement on a noisy qubit results in the noise expansion to its neighbors with a phase. From Eq.~\eqref{eq:pauli:y:a}, we get the results presented in Table~\ref{tab:pauli:y}, which show the transformations of the elements of the noise map of the measured qubit $a$. Moreover, in Fig.~\ref{fig:manipulation:y} we graphically show how the different Pauli noise operators on a qubit change after the local Pauli $Y$ measurement on it. For any qubit $b\in N_a$, the elements of their noise maps change following Eq.~\eqref{eq:kraus:in:Na}. For any qubits $j\neq a$ and $j\notin N_a$, the elements of their noise maps do not change as described in Eq.~\eqref{eq:kraus:no:change}. Note that the neighborhood of these qubits $j$ does not change after a local Pauli $Y$ measurement on $a$. In general, the noise operators now act on the graph state corresponding to the resulting graph $G'$ after the measurement.

\begin{table}[ht]
\centering
\begin{tabular}{|clc|}
\hline
Elements of $\mathcal{M}_a \rho$ &               & Elements of $\widetilde{\mathcal{M}}_a \rho'$ \\ \hline \hline
$\rho $         & $\rightarrow$ &  $\rho'$           \\
$X_a \rho X_a = Z_{N_a} \rho Z_{N_a}$         & $\rightarrow$ &  $Z_{N_a} \rho' Z_{N_a}$           \\
$Y_a \rho Y_a = Z_a Z_{N_a} \rho Z_a Z_{N_a}$         & $\rightarrow$ & $\rho'$           \\
$Z_a \rho Z_a$         & $\rightarrow$ & $Z_{N_a} \rho' Z_{N_a}$           \\ \hline
\end{tabular}
\caption{Elements of the noise map of $a$ before, $\mathcal{M}_a\rho$, and after, $\widetilde{\mathcal{M}}_a\rho'=L_{y,\pm}^{(a)}(\mathcal{M}_{a}\rho)(L_{y,\pm}^{(a)})^{\dagger}$, a local Pauli $Y$ measurement on $a$.}
\label{tab:pauli:y}
\end{table}

Notably, the depolarizing maps for $j \neq a$ will be the same as if we would write them after the measurement. However, the depolarizing noise map for $a$ results in Eq.~\eqref{eq:depolar:map:a}.

As stated in Sec.~\ref{ssec:graph:states}, when performing several consecutive Pauli measurements the local unitaries have to be taken into account. Because the local unitaries of $U_{y,\pm}^{(a)}$ and the local noise operator $Z_a$ do not commute, the noise spreads to the neighborhood of $a$ changing the noise pattern. Then if a following $Y$ measurement is performed on any of the neighborhood qubits, this is applied on top of some additional noise (Pauli $Z$) which, once again, does not commute with the local unitary. Therefore, we conclude that in a noisy graph state the order in which consecutive local Pauli $Y$ measurements are performed matters. As a consequence, different patterns of measurements will lead to different noise patterns.  

\subsubsection{Pauli X measurement}\label{a:nsf:x:measurement}
\begin{figure}[!ht]
    \centering
    \includegraphics[width=0.5\columnwidth]{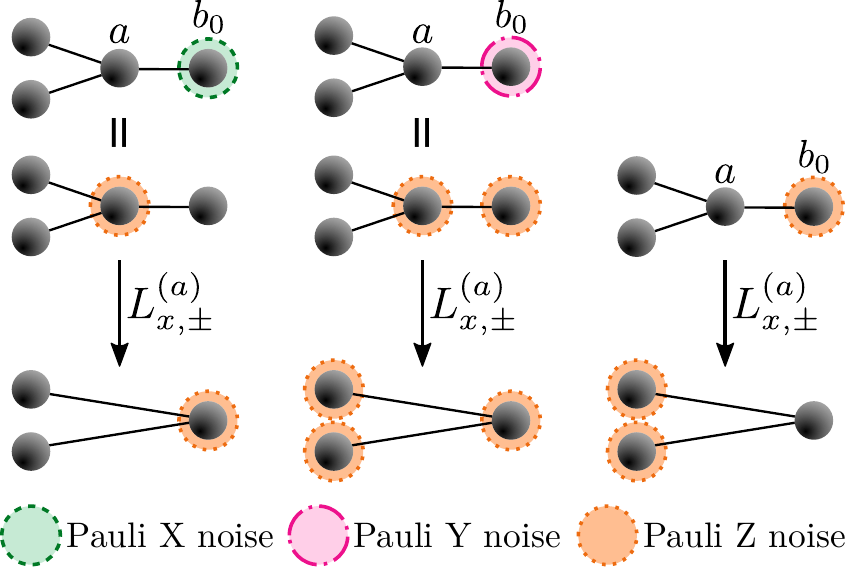}
    \caption{Graphical representation of a local Pauli measurement in the $X$ basis on a noiseless qubit but the special qubit $b_0$ is subject to Pauli noise, $X$, $Y$ or $Z$.}
    \label{fig:manipulation:x:b0}
\end{figure}

The noise operator $Z_j$ commutes with $L_{x, \pm}^{(a)}$ for $j\neq a, b_0$. However, the noise operator $Z_a$ does not, as
\begin{equation}\label{eq:pauli:x:a}
    L_{x, \pm}^{(a)} Z_a = Z_{b_0} Z_{N_{b_0}} L_{x, \pm}^{(a)}.
\end{equation}
Similarly, the noise operator $Z_{b_0}$ does not commute either
\begin{equation}\label{eq:pauli:x:b0}
    L_{x, \pm}^{(a)} Z_{b_0} = Z_{N'_{b_0}} L_{x, \pm}^{(a)},
\end{equation}
where $N'_{b_0}$ denotes the neighborhood of $b_0$ after the measurement ($N'_{b_0}=N_a \setminus \{b_0\}$). From Eqs.~\eqref{eq:pauli:x:a} and \eqref{eq:pauli:x:b0} we get the results of Tables~\ref{tab:pauli:x:a} and 
\ref{tab:pauli:x:b0}, which show the transformations of the elements of the noise map of the measured qubit $a$ and of $b_0$ correspondingly. Moreover, in Fig.~\ref{fig:manipulation:x:a} and Fig.~\ref{fig:manipulation:x:b0}, we graphically show how the different Pauli noise operators of $a$ and $b_0$ change after the local Pauli $X$ measurement on $a$, correspondingly. For any other qubit $j\neq a, b_0$, the elements of their noise maps change following Eq.~\eqref{eq:kraus:not:a}, where $N'_j$ denotes the neighborhood after the Pauli $X$ measurement. 

\begin{table}[ht]
\centering
\begin{tabular}{|clc|}
\hline
Elements of $\mathcal{M}_a \rho$ &               & Elements of $\widetilde{\mathcal{M}}_a \rho'$ \\ \hline \hline
$\rho $         & $\rightarrow$ &  $\rho'$           \\
$X_a\rho X_a = Z_{N_a} \rho Z_{N_a}$         & $\rightarrow$ &  $\rho'$           \\
$Y_a\rho Y_a = Z_a Z_{N_a} \rho Z_a Z_{N_a}$         & $\rightarrow$ & $Z_{b_0} Z_{N_{b_0}} \rho' Z_{b_0} Z_{N_{b_0}}$           \\
$Z_a \rho Z_a$         & $\rightarrow$ & $Z_{b_0} Z_{N_{b_0}} \rho' Z_{b_0} Z_{N_{b_0}}$           \\ \hline
\end{tabular}
\caption{Elements of the noise map of $a$ before, $\mathcal{M}_a\rho$, and after, $\widetilde{\mathcal{M}}_a\rho'=L_{x,\pm}^{(a)}(\mathcal{M}_{a}\rho)(L_{x,\pm}^{(a)})^{\dagger}$, a local Pauli $X$ measurement on $a$.}
\label{tab:pauli:x:a}
\end{table}
\begin{table}[ht]
\centering
\begin{tabular}{|clc|}
\hline
Elements of $\mathcal{M}_{b_0} \rho$ &               & Elements of $\widetilde{\mathcal{M}}_{b_0} \rho'$ \\ \hline \hline
$\rho $         & $\rightarrow$ &  $\rho'$           \\
$X_{b_0}\rho X_{b_0} = Z_{N_{b_0}} \rho Z_{N_{b_0}}$         & $\rightarrow$ &  $Z_{b_0} \rho' Z_{b_0}$           \\
$Y_{b_0}\rho Y_{b_0} = Z_{b_0} Z_{N_{b_0}} \rho Z_{b_0} Z_{N_{b_0}}$ & $\rightarrow$ & $Z_{b_0} Z_{N'_{b_0}} \rho' Z_{b_0} Z_{N'_{b_0}}$  \\
$Z_{b_0} \rho Z_{b_0}$         & $\rightarrow$ & $Z_{N'_{b_0}} \rho' Z_{N'_{b_0}}$           \\ \hline
\end{tabular}
\caption{Elements of the noise map of $b_0$ before, $\mathcal{M}_{b_0}\rho$, and after, $\widetilde{\mathcal{M}}_{b_0}\rho' = L_{x,\pm}^{(a)} (\mathcal{M}_{b_0}\rho) (L_{x,\pm}^{(a)})^{\dagger}$, a local Pauli $X$ measurement on $a$.}
\label{tab:pauli:x:b0}
\end{table}

Notably, the depolarizing maps for $j \neq a$ will be the same as if we would write them after the measurement. However, the depolarizing noise map for $a$ results in 
\begin{equation}\label{eq:map:x:meas}
    \widetilde{\mathcal{M}}_a\rho' =p\rho'+\frac{1-p}{2}\left(\rho'+ Z_{b_0} Z_{N_{b_0}}\rho'Z_{b_0}Z_{N_{b_0}}\right).
\end{equation}

As for the local Pauli $Y$ measurement, the order of consecutive $X$ measurements matters, as the noise will not spread in the same manner depending on the order. This also holds for a combination of Pauli measurements in the $X$ and $Y$ basis.

\subsection{Merging}\label{a:nsf:merging}
\begin{figure}[ht]
    \centering
    \includegraphics[width=0.66\columnwidth]{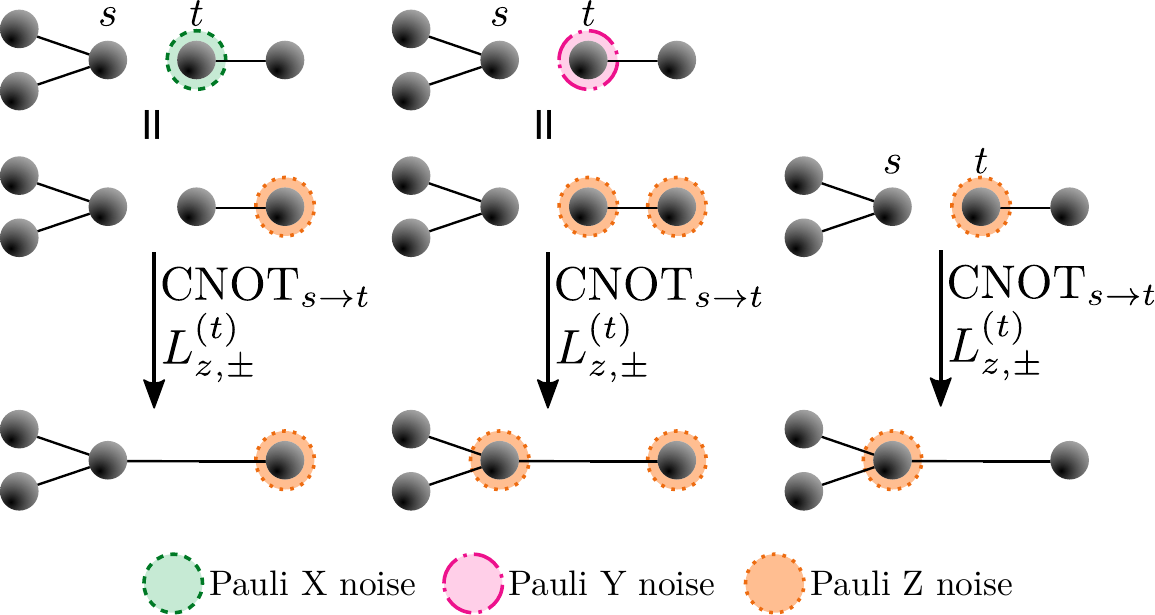}
    \caption{Graphical representation of a merge operation between two graphs, where the target qubit is subject to Pauli noise, $X$, $Y$ or $Z$.}
    \label{fig:manipulation:merge}
\end{figure}

As described in Sec.~\ref{sssec:merging}, the merging of two graphs is done via a CNOT gate between a source and a target and a $Z$ measurement on the target. Let us first analyze the entangling operation. The noise operator $Z_j$ commutes with $\text{CNOT}_{s\rightarrow t}$ for $j\neq t$. However, the noise operator $Z_t$ does not, such that 
\begin{equation}
\text{CNOT}_{s\rightarrow t} Z_t = Z_s Z_t\text{CNOT}_{s\rightarrow t}.
\end{equation}
Thus, $Z_t$ noise spreads to the source qubit. Now, the target qubit is measured in the $Z$ basis; using Eq.~\eqref{eq:pauli:z:a} we get
\begin{equation}\label{eq:merge:t}
    L_{z,\pm}^{(t)}\text{CNOT}_{s\rightarrow t}Z_t = \pm Z_s L_{z,\pm}^{(t)}\text{CNOT}_{s\rightarrow t}.
\end{equation}
From Eq.~\eqref{eq:merge:t} we get the results in Table~\ref{tab:merge:t}, which show the transformation of the elements of the noise map of the target qubit $t$. Moreover, in Fig.~\ref{fig:manipulation:merge}, we graphically show how the different Pauli noise operators of $t$ change after the merging. Importantly, the noise operators for $s$ are not altered after the manipulation. For any qubit such that $j\neq s, t$, the elements of their noise maps change following Eq.~\eqref{eq:kraus:not:a}, where $N'_j$ denotes the neighborhood after the merging operation.
\begin{table}[ht]
\centering
\begin{tabular}{|clc|}
\hline
Elements of $\mathcal{M}_{t} \rho$ &               & Elements of $\widetilde{\mathcal{M}}_{t} \rho'$ \\ \hline \hline
$\rho $         & $\rightarrow$ &  $\rho'$           \\
$X_t\rho X_t = Z_{N_{t}} \rho Z_{N_{t}}$         & $\rightarrow$ &  $Z_{N_t} \rho' Z_{N_t}$           \\
$Y_t \rho Y_t = Z_{t} Z_{N_{t}} \rho Z_{t} Z_{N_{t}}$ & $\rightarrow$ & $Z_{s} Z_{N_{t}} \rho' Z_{s} Z_{N_{t}}$  \\
$Z_{t} \rho Z_{t}$         & $\rightarrow$ & $Z_{s} \rho' Z_{s}$           \\ \hline
\end{tabular}
\caption{Elements of the noise map of $t$ before, $\mathcal{M}_{t}\rho$, and after, $\widetilde{\mathcal{M}}_{t}\rho' = L_{z,\pm}^{(t)} \text{CNOT}_{s\rightarrow t} (\mathcal{M}_{t}\rho) (L_{z,\pm}^{(t)})^{\dagger} (\text{CNOT}_{s\rightarrow t})^{\dagger}$, a merging between two graph states using qubits $s$ and $t$.}
\label{tab:merge:t}
\end{table}

Notably, the depolarizing maps for $j \neq t,s$ will be the same as if we would write them after the merging. However, the depolarizing noise map for $t$ results in 
\begin{equation}
    \widetilde{\mathcal{M}}_t\rho' = p\rho' + \frac{1-p}{4} (\rho' + Z_s\rho'Z_s  + Z_{N_t}\rho' Z_{N_t} + Z_s Z_{N_t}\rho'Z_s Z_{N_t}), 
\end{equation}
and the depolarizing map for $s$ is not altered.

As described in Sec.~\ref{sssec:merging}, the full merging of two graphs is done via a CNOT gate between a source and a target, a $Z$ measurement on the target and a $Y$ measurement on the source. To analyze the noise of this operation one just needs to use the already studied results for the merging and the Pauli $Y$ measurement. Thus, the noise operators of $t$ change as
\begin{equation}
    Z_t^{\alpha} Z_{N_t}^{\beta} \rightarrow Z_{N_s}^{\alpha} Z_{N_t}^{\alpha + \beta},
\end{equation}
and for $s$,
\begin{equation}
    Z_s^{\alpha} Z_{N_s}^{\beta} \rightarrow Z_{N_t}^{\alpha} Z_{N_s}^{\alpha + \beta}.
\end{equation}
For any qubit $b$ in the neighborhood of $s$ after the simple merging operation, $b \in (N_s\cup N_t)\setminus (N_s\cap N_t) \setminus \{s\}\setminus \{t\}$, the elements of their noise maps change following Eq.~\eqref{eq:kraus:in:Na}. For any other qubit, the elements of their noise maps do not change as described in Eq.~\eqref{eq:kraus:no:change}. 

Notably, the depolarizing maps for $j \neq t,s$ will be the same as if we would write them after the merging. However, the depolarizing noise map for $t$ and $s$ results in 
\begin{equation}
    \widetilde{\mathcal{M}}_t\rho'=\widetilde{\mathcal{M}}_s\rho'= p\rho' + \frac{1-p}{4} (\rho' + Z_{N_s}\rho'Z_{N_s}  + Z_{N_t}\rho' Z_{N_t} + Z_{N_s} Z_{N_t}\rho'Z_{N_s} Z_{N_t}).
\end{equation}

\section{Remarks on general noise maps}\label{a:remarks:general:noise}
Here we comment on general noise maps and the limitations of the noisy stabilizer formalism. A general multiqubit noise channel acting on systems $a_1a_2 \cdots a_k$ can always be written in the Pauli basis, i.e., is of the form 
\begin{equation}
    \mathcal{E}_{a_1a_2 \cdots a_k} \rho = \sum_{i,j=1}^{4^k} \lambda_{ij} N_i \rho N_j^\dagger,
\end{equation}
where the noise operators are given by tensor products of Pauli operators, $N_i= \sigma_{i_1}^{(a_1)}\otimes\sigma_{i_2}^{(a_2)} \otimes \cdots \otimes \sigma_{i_k}^{(a_k)}$. 

While we have shown that for diagonal Pauli channels acting on graph states, a simple replacement of Pauli operators by commuting Pauli $Z$ operators suffices, the situation is not so straightforward for arbitrary channels including off-diagonal elements. In this case, phases $\pm i$ that appear in the action of $Y$ on a graph state and additional ones that appear due to the commutation relation of Pauli operators, need to be considered. 

Let us first consider phases due to the action of the noise operator $Y_a$. While $Y_a|G\rangle = -i Z_a Z_{N_a}|G\rangle$, we have that $\langle G|Y_a = i \langle G|Z_a Z_{N_a}$. These phases cancel for any Pauli channel that includes only diagonal terms when written in the Pauli basis, however, phases in off-diagonal terms remain, e.g., $Z_a|G\rangle\langle G|Y_a = i Z_a|G\rangle \langle G|Z_a Z_{N_a}$. There are, however, additional phases appearing, which originate from the Pauli commutation relations, when Pauli operators originating from the action of noise on different qubits act on the same one. These phases have to be taken into account and computed for each tensor product of Pauli operators separately. Hence a simple replacement and update of individual noise maps to commuting $Z$-noise maps are not sufficient in this case, making the treatment of such noise channels inefficient.

We illustrate this with a simple example. Consider a graph state of two qubits 1 and 2 that are connected by an edge $(1,2)$, that is, $N_1=2$ and $N_2=1$. It follows that $Y_1\otimes Y_2|G\rangle=(\mathbbm{1}\otimes Y_2)(Y_1 \otimes \mathbbm{1})|G\rangle = (\mathbbm{1}\otimes Y_2)(-i)(Z_1\otimes Z_2)|G\rangle = (-1)(-i)(Z_1 \otimes Z_2)(\mathbbm{1} \otimes Y_2)|G\rangle = (-1)(-i)(-i)|G\rangle = |G\rangle$. The additional $(-1)$ factor compared to simply replacing each of the $Y$ operators by $(-i)Z_1\otimes Z_2$ comes from $Y_2Z_2 = - Z_2Y_2$. For the diagonal term $Y_1\otimes Y_2 |G\rangle\langle G| Y_1\otimes Y_2 = |G\rangle \langle G|$, the phase does not matter. However, the phase is relevant for off-diagonal elements. Such a situation appears also in the case when independent noise maps that include off-diagonal terms act on different qubits. The problem is that one needs to consider all (exponentially many) combinations of noise operators from different maps, and not just maps independently.

We however stress again that this is not a severe restriction of our method. On the one hand, many relevant noise models are of Pauli diagonal form, including uncorrelated or correlated dephasing, bit-flip noise and depolarizing noise (white noise). On the other hand, one can enforce noise to be of such Pauli diagonal form via the application of random unitary operations before and after the action of the noise channel (see \cite{dur_standard}), without changing the diagonal elements and the Choi-Jamio\l{}okowsky fidelity of the map.

\section{Strategy - side to side}\label{a:side:to:side}
In Sec.~\ref{sec:application} we analyzed the manipulation of an $N$-qubit 1D cluster, where each qubit is subject to depolarizing noise, to a two-qubit graph state between the end qubits. Here we show the proof of the weight vector in Eq.~\eqref{eq:side:to:side} that describes the noise pattern for the strategy of measuring side to side. 

It is important to note that depending on which direction the measurements are performed, the noise accumulates on one of the ends of the cluster. For example, if the measurements are performed from 1 to $N$, the noise accumulates more on $N$, meaning that $w_{\alpha}=0$. However, as proved before, the fidelity is not affected by this. Moreover, the proof is done for the direction $1\rightarrow N$. The proof for the opposite direction is completely analogous to the one presented here. 

Before stating the proof, we introduce some additional notation. In this manipulation, $n=N-2$ measurements are performed and $s$ denotes which measurement step has been done, which is the Pauli $Y$ measurement of qubit $s+1$. Then $\rho^{(s)}$ denotes the graph state after the $s$th measurement and $\mathcal{M}_m^{(s)}$ denotes the noise map of qubit $m$ after the $s$th measurement.

\subsection{Noise map of qubit 2}
First, we will see how the noise map for qubit 2 evolves after each measurement. This map after the $Y$ measurement of qubit 2 can be computed using Eq.~\eqref{eq:depolar:map} and Table~\ref{tab:commutations} and is
\begin{equation}
    \mathcal{M}_2^{(1)}\rho^{(1)} = p\rho^{(1)} + \frac{1-p}{2}\left(\rho^{(1)}+Z_1Z_3 \rho^{(1)}Z_1Z_3 \right).
\end{equation}
Then we perform the second measurement, a Pauli $Y$ measurement on qubit 3. Now, the noise map of qubit 2 is updated to
\begin{align}
    \mathcal{M}_2^{(2)}\rho^{(2)} & = p\rho^{(2)} + \frac{1-p}{2}\left(\rho^{(2)}+Z_4 \rho^{(2)}Z_4 \right). 
\end{align}
We assume the following for any $s$:
\begin{equation}
\begin{aligned}
\mathcal{M}_2^{(s)}\rho^{(s)} =
\begin{cases}
    p\rho^{(s)} + \frac{1-p}{2}\left(\rho^{(s)}+Z_{s+2} \rho^{(s)}Z_{s+2} \right) \text{ for } s \text{ even} \\
    p\rho^{(s)} + \frac{1-p}{2}\left(\rho^{(s)}+Z_1Z_{s+2} \rho^{(s)}Z_1Z_{s+2} \right) \text{ for } s \text{ odd}.
\end{cases}
\end{aligned}
\end{equation}
Now, we want to prove that these assumptions hold for $s+1$, using Table~\ref{tab:commutations},
\begin{equation}
\mathcal{M}_2^{(s+1)}\rho^{(s+1)}  =
\begin{cases}
    p\rho^{(s+1)} + \frac{1-p}{2}\left(\rho^{(s+1)}+Z_{s+3} \rho^{(s+1)}Z_{s+3} \right) \text{ for } s \text{ even} \\
    p\rho^{(s+1)} + \frac{1-p}{2}\left(\rho^{(s+1)}+Z_1Z_{s+3} \rho^{(s+1)}Z_1Z_{s+3} \right) \text{ for } s \text{ odd}.
\end{cases}
\end{equation}
Therefore, after the last measurement $s=n$,
\begin{equation}\label{eq:map:2}
\mathcal{M}_2^{(n)}\rho^{(n)} = 
    \begin{cases}
        \mathcal{M}_{\gamma}\rho' \text{ for } n \text{ even} \\
        \mathcal{M}_{\alpha}\rho' \text{ for } n \text{ odd}.
    \end{cases}
\end{equation}

\subsection{Noise maps of all qubits}
Following this structure we can prove the same for any other qubit $m$, such that $m=2, \dots, N-1$. If $m$ is even, its final noise map will be the same as for $m=2$ presented in Eq.~\eqref{eq:map:2}. However, if $m$ is odd, its corresponding noise map will be 
\begin{equation}
\mathcal{M}_m^{(n)}\rho^{(n)}=
    \begin{cases}
        \mathcal{M}_{\alpha}\rho' \text{ for } n \text{ even} \\
        \mathcal{M}_{\gamma}\rho' \text{ for } n \text{ odd}.
    \end{cases}
\end{equation}

Therefore, if $n$ is even, $w_{\gamma}=n/2$ and $w_{\alpha}= n/2$. Moreover, if $n$ is odd, $w_{\gamma}=(n-1)/2$ and $w_{\alpha}=(n+1)/2$. This proves the weight vector in Eq.~\eqref{eq:side:to:side}.

\section{Exact fidelities}\label{a:fidelity}
In this appendix, we provide the exact formulas for the fidelity of the final two-qubit graph state that results from manipulating a 1D cluster, which corresponds to the example described in Sec.~\ref{sec:application}, via different strategies. We compute the fidelities using the weight vectors of each strategy presented in Sec.~\ref{ssec:strategies} and the general formula for the fidelity for this specific scenario in Eq.~\eqref{eq:general:fidelity}. Recall that for this example $t=0$, the size of the initial cluster is $N$ qubits, and $n=N-2$ of them are measured in the $Y$ basis. To have a formula per strategy we make use of the functions
\begin{equation}
    g({x})=\frac{1-(-1)^{x}}{2}, \quad
    f(x)=\frac{1+(-1)^{x}}{2}.
\end{equation}
For the side-to-side strategy (sts), the exact fidelity is
\begin{equation}
    F_{\text{sts}}(n,p)=\frac{1}{4} \left(1+ p^{2+\frac{n+g(n)}{2}}+p^{2+\frac{n-g(n)}{2}}+p^{n+2}\right).
\end{equation}
For the every-second-qubit strategy (esq), the exact fidelity for $n=3l+\beta$, where $l\in \mathbb{Z}^+$ and $\beta=0, \, 1, \, 2$, is
\begin{equation}
    F_{\text{esq}}(l,\beta,p)=\frac{1}{4}\left[1 + p^{2 l+2}\left(p^{\frac{\beta-g(\beta)}{2}}+ p^{\frac{\beta + g(\beta)}{2}}+ p^{\beta} \right)\right].
\end{equation}
For the pairs strategy (pair), the exact fidelity for $n=4l+\beta$, where $l\in \mathbb{Z}^+$ and $\beta=0, \, 1, \, 2, \, 3$, is
\begin{equation}
    F_{\text{pair}}(l,\beta,p)=\frac{1}{4}\left[1 + p^{2 l+2}\left(p^{f(\beta)\frac{\beta-1}{2}}+ p^{l+\beta-1-f(\beta)\frac{\beta-1}{2}} + p^{l+\beta-1} \right)\right].
\end{equation}
\end{document}